\documentclass[journal,draftclsnofoot,onecolumn,12pt,comsoc, twoside,a4paper]{IEEEtran}

\input{packages_commands_theorems.tex}

\usepackage[noamssymbols]{newtxmath}
\usepackage{eucal}

\usepackage[ruled,vlined,linesnumbered]{algorithm2e}
\SetKwInput{KwIn}{Initialization}

\usepackage[capitalise]{cleveref}

\usepackage{optidef}

\usepackage{bm}

\newcommand{\diag}{\mathop{\mathrm{diag}}}

\newcommand*\diff{\mathop{}\!\mathrm{d}}

\usepackage{enumitem}

\usepackage{acronym}
\acrodef{CSI}[CSI]{channel state information}
\acrodef{RIS}[RIS]{reconfigurable intelligent surface}
\acrodef{SNR}[SNR]{signal-to-noise ratio}
\acrodef{SCD}[SCD]{\emph{successive cancellation decoding}}
\acrodef{iid}[i.i.d.]{independent and identically distributed}
\acrodef{BS}[BS]{base station}
\acrodef{PDF}[PDF]{probability density function}
\acrodef{CDF}[CDF]{cumulative distribution function}
\acrodef{CCDF}[CCDF]{complementary cumulative distribution function}
\acrodef{VaR}[VaR]{\emph{value-at-risk}}
\acrodef{CVaR}[CVaR]{\emph{conditional value-at-risk}}
\acrodef{AWGN}[AWGN]{additive white Gaussian noise}
\acrodef{WLOG}[w.l.o.g.]{without loss of generality}
\acrodef{MAML}[MAML]{model-agnostic meta-learning}
\acrodef{LDM}[LDM]{layered division multiplexing}
\acrodef{LDM_cap}[LDM]{Layered division multiplexing}
\acrodef{BSMs}[BSMs]{basic safety messages}
\acrodef{V2X}[V2X]{vehicle-to-everything}
\acrodef{QoS}[QoS]{quality-of-service}
\acrodef{ATSC}[ATSC 3.0]{Advanced Television Systems Committee}
\acrodef{MTC}[MTC]{Machine-Type Communication}

\begin{document}
	
\bstctlcite{IEEEexample:BSTcontrol} 

\title{Learning to Broadcast for Ultra-Reliable Communication with Differential Quality of Service via the Conditional Value at Risk}

\author{Roy~Karasik,~\IEEEmembership{Graduate Student Member,~IEEE,}
	Osvaldo~Simeone,~\IEEEmembership{Fellow,~IEEE,}
	Hyeryung~Jang,~\IEEEmembership{Member,~IEEE,}
	and~Shlomo~Shamai~(Shitz),~\IEEEmembership{Life Fellow,~IEEE}
	\thanks{This work has been supported by the European Research Council (ERC) under the European Union’s Horizon 2020 Research and Innovation Programme (Grant Agreement Nos. 694630 and 725731).}
	\thanks{Roy Karasik and Shlomo Shamai (Shitz) are with the Department of Electrical and Computer Engineering, Technion---Israel Institute of Technology, Haifa 32000, Israel (e-mail: royk@campus.technion.ac.il; sshlomo@ee.technion.ac.il).}
	\thanks{Osvaldo Simeone is with King's Communications, Learning \& Information Processing (KCLIP) lab, Centre for Telecommunications Research, Department of Engineering, King’s College London, London WC2R 2LS, U.K. (e-mail: osvaldo.simeone@kcl.ac.uk).}
	\thanks{Hyeryung Jang is with the Department of Artificial Intelligence, Dongguk University, Seoul 04620, South Korea (e-mail: hyeryung.jang@dgu.ac.kr).}
}

\maketitle

\begin{abstract}
	Broadcast/multicast communication systems are typically designed to optimize the outage rate criterion, which neglects the performance of the fraction of clients with the worst channel conditions. Targeting ultra-reliable communication scenarios, this paper takes a complementary approach by introducing the \ac{CVaR} rate as the expected rate of a worst-case fraction of clients. To support differential \ac{QoS} levels in this class of clients, \ac{LDM} is applied, which enables decoding at different rates. Focusing on a practical scenario in which the transmitter does not know the fading distribution, layer allocation is optimized based on a dataset sampled during deployment. The optimality gap caused by the availability of limited data is bounded via a generalization analysis, and the sample complexity is shown to increase as the designated fraction of worst-case clients decreases. Considering this theoretical result, meta-learning is introduced as a means to reduce sample complexity by leveraging data from previous deployments. Numerical experiments demonstrate that \ac{LDM} improves spectral efficiency even for small datasets; that, for sufficiently large datasets, the proposed mirror-descent-based layer optimization scheme achieves a \ac{CVaR} rate close to that achieved when the transmitter knows the fading distribution; and that meta-learning can significantly reduce data requirements.
\end{abstract}


\section{Introduction}
Layered division multiplexing (LDM) has been introduced in several standards as an effective means to support differential \acf{QoS} in broadcast and multicast services. With \ac{LDM}, multiple independent sub-messages, or layers, are superimposed, enabling the decoding of a different number of messages depending on the channel conditions, thus supporting communication at a variable rate \cite{tajer2021broadcast,barquero2015LDM,verdu2010variable,zhang2016layered}. The most common use of \ac{LDM} is for multimedia broadcast, as adopted by the \ac{ATSC} \cite{zhang2016layered,park2916low}, in which \ac{LDM} supports a robust configuration for mobile receivers and a high-capacity connection for fixed receivers. Other applications include \ac{MTC} and Industry 4.0, in which \ac{LDM} is considered as a tool to deliver critical control services and best-effort monitoring services \cite{arruti2017qos,arruti2017unequal,montalban2020noma}; as well as \ac{V2X} communications \cite{boban2018connected}. In \ac{V2X} systems, traffic light status and pedestrian detection information in intelligent intersections \cite{kostopoulos2020use} can be broadcast together with high-definition map transmission for enhanced autonomous driving \cite{wang2019cooperative}.

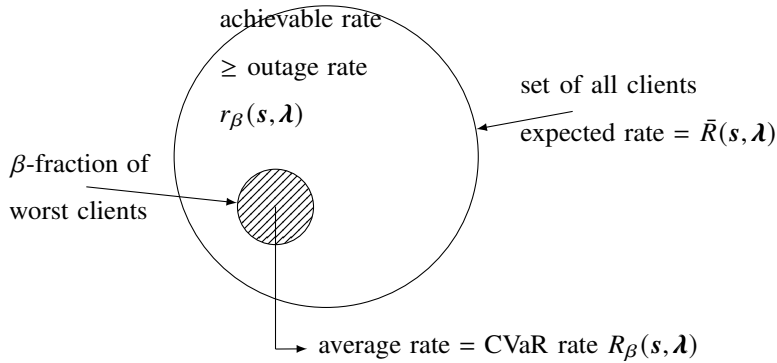
\begin{figure}[!t]
	\centering
	\begin{tikzpicture}[>=latex]
	\def\setRadius{2cm}
	\def\subsetRadius{0.5cm}
	
	\coordinate (setCenter) at (0, 0);
	\coordinate (subsetCenter) at ($(setCenter) - (\setRadius/3,\setRadius/3)$);

	\node[circle, draw=black,minimum size=2*\setRadius] (circ) at (setCenter){};
	\node[circle, draw=black, pattern=north east lines, minimum size=2*\subsetRadius] (subCirc) at (subsetCenter) {};
	\node [font=\small, above = 0.2cm of circ.center, text width=1.4*\setRadius] {achievable rate $\geq$ outage rate $r_{\beta}(\bm{s},\pmb{\lambda})$};
	\node (A) [font=\small, right = 0.4cm of circ.east, yshift=0.6cm, align=left] {set of all clients \\ expected rate $=$ $\bar{R}(\bm{s},\pmb{\lambda})$};
	\draw[->] ($(A.center)-(1cm,0)$) -- (circ);
	\node (B) [font=\small, left = 0.0cm of circ.west, yshift=-0.4cm, text width = 2cm] {$\beta$-fraction of worst clients};
	\draw[->] (B.center) -- (subCirc);
	\node (C) [font=\small, below = 0.2cm of circ.south, xshift=1.2*\setRadius] {average rate $=$ CVaR rate $R_\beta(\bm{s},\pmb{\lambda})$};
	\node (D) [below = 1.23cm of subCirc.south] {};
	\draw (subCirc.center) -- (D.center);
	\draw[->] (D.center) -- (C);
\end{tikzpicture}
	\caption{Illustration of the standard $\beta$-outage rate and of the proposed $\beta$-\acs{CVaR} rate.}
	\label{fig:cvar}
\end{figure}

In many of the use cases of broadcast and multicast services, it is essential to ensure that a large fraction of clients receives at least some of the information being transmitted, such as \ac{BSMs} in \ac{V2X} \cite{liu2016delivering_short}. The corresponding standard design objective is the transmission of a single message at the outage rate \cite{angjelichinoski2019statistical,jurdi2018variable,popovski2019wireless,hampel2019ultra}. For outage probability $\beta\in[0,1]$, the $\beta$-outage rate is the largest rate that can be guaranteed with probability $1-\beta$, i.e., that can be achieved by a fraction $1-\beta$ of all possible clients (see \cref{fig:cvar}). Transmitting at the $\beta$-outage rate implies that the fraction $\beta$ of clients with the worst channel conditions cannot decode the (single) message being broadcast. 

Targeting the design of \ac{LDM} broadcasting/multicasting in ultra-reliable communication systems, this work takes the complementary approach of focusing on the differential \ac{QoS} performance of the $\beta$-fraction of clients with the worst instantaneous channel conditions (see \cref{fig:cvar}). To this end, we introduce the $\beta$-\acf{CVaR} rate as the expected rate achieved by the $\beta$-fraction of clients with the worst fading channels via \ac{LDM}. Unlike the $\beta$-outage rate criterion, the $\beta$-\ac{CVaR} rate is concerned with the performance of the worst-case $\beta$-fraction of clients, enabling clients in this class to decode at different rates. 

Maximizing the $\beta$-\ac{CVaR} rate requires adjusting the layers’ rates and power levels as a function of the channel distribution \cite{shamai2003broadcast}. However, in practice, this distribution is unknown. Accordingly, in this paper, we assume the transmitter has access to a dataset sampled during deployment, from which the rate and power allocation for each layer are optimized. We explore theoretic and algorithmic aspects of this design problem, including also extensions to learning to learn, or meta-learning \cite{thrun2012learning}.

\emph{Related Work:} \ac{LDM}, also known as the \emph{broadcast approach}, has been extensively studied as means to improve spectral efficiency in various scenarios. A comprehensive survey of the state-of-the-art is available in \cite{tajer2021broadcast}, and we mention here some representative examples. The broadcast approach for slowly fading single-user channels was investigated in \cite{shamai2003broadcast}, where it was shown that transmitting multiple layers can increase the expected achievable rate. The gain of the broadcast approach was also demonstrated in \cite{liu2002optimal} for finite number of layers. Specifically, for quasi-static Rayleigh fading channel, two layers were shown to achieve most of the throughput gain. Importantly, unlike our work, both references \cite{shamai2003broadcast} and \cite{liu2002optimal} assume that the transmitter knows the fading distribution. 

With respect to ultra-reliable communication, multicast beamforming was studied in \cite{ntranos2009multicast} with the goal of minimizing the outage probability, and an approximate solution was obtained for a Gaussian mixture channel with up to three Gaussian kernels. For unknown fading distribution, several gradient-based algorithms were proposed in \cite{shi2018learning} to optimize beamforming based on a dataset of channel samples. Similarly, an alternating gradient descent algorithm was recently proposed in \cite{fang2020outage} for the joint optimization of the precoding weights and the \ac{RIS} reflection pattern in \ac{RIS}-aided communication system.

Optimization of the \ac{CVaR} statistic was first introduced in \cite{rockafellar2000optimization} for financial applications. Since then, it has been considered in a variety of fields \cite{filippi2020conditional}. For communication systems, the \ac{CVaR} statistic was applied in \cite{alsenwi2019embb} for risk-sensitive resource allocation scheme targeting low-latency traffic; in \cite{li2019statistical} for statistical \ac{QoS} estimation in a shared spectrum; and in \cite{wu2020energy} for robust computation offloading from mobile devices to infrastructure nodes.

A review of meta-learning with emphasis on applications to communication systems is available in \cite{simeone2020from}. Representative examples include meta-learning for learning to demodulate  \cite{park2021learning,cohen2021learning} or decode \cite{jiang2019mind}; for end-to-end learning of encoder and decoder \cite{park2020meta}; for beamforming adaptation \cite{yuan2021transfer,zhang2021embedding}; for proactive resource allocation \cite{yuan2021meta,park2021predicting}; and for channel estimation \cite{mao2019roemnet}.

\emph{Main Contributions:} This work introduces and studies the concept of $\beta$-\ac{CVaR} rate for the problem of \ac{LDM}-based broadcasting/multicasting in systems with a single-antenna \ac{BS} serving single-antenna clients. The channel coefficients and the fading distribution are assumed to be unknown to the \ac{BS}, which optimizes layer allocation based on a dataset sampled during deployment. We address both information-theoretic and algorithmic aspects, with the specific contributions being as follows.
\begin{itemize}[leftmargin=*]
	\item We introduce the concept of $\beta$-\ac{CVaR} rate as the average rate obtained for the $\beta$-fraction of clients with the worst channel conditions (see \cref{fig:cvar}). This novel criterion targets ultra-reliable broadcasting/multicasting applications, while allowing for differential worst-case \ac{QoS} guarantees.
	\item As a special case of the problem of maximizing the $\beta$-\ac{CVaR} metric, we review the optimization of power and rate allocation for the expected achievable rate when the fading distribution is known and infinite layers are applied. For this scenario, we bound the optimality gap caused by the availability of limited data via a generalization analysis \cite{lee2020learning}.
	\item Moving beyond the expected rate metric, we address the problem of maximizing the $\beta$-\ac{CVaR} rate. At a theoretical level, we characterize the number of samples required to maintain a desired optimality gap, showing that the sample complexity increases as the fraction $\beta$ decreases. At an algorithmic level, we introduce a mirror-descent based scheme \cite{beck2003mirror} to maximize an empirical estimate of the $\beta$-\ac{CVaR} rate. 
	\item In light of our theoretical result that the sample complexity increases as $\beta^{-1}$ as $\beta$ decreases, we address the problem of reducing sample complexity via meta-learning. By leveraging data from multiple previous deployments, each with different fading distributions, meta-learning aims at decreasing data requirements on new deployments.
	\item Numerical results demonstrate that broadcasting multiple layers improves spectral efficiency even for small datasets, and that, for sufficiently large dataset, the expected rate and $\beta$-\ac{CVaR} rate are close to that achieved when the \ac{BS} knows the fading distribution, confirming the sample complexity analysis. In addition, meta-learning is shown to be effective in decreasing the sample complexity for low outage probabilities.
\end{itemize}

\emph{Organization:} The rest of the paper is organized as follows. In \cref{sec:model}, we present an information-theoretic model for a multi-layer broadcast channel with no \ac{CSI}. Maximization of the expected achievable rate is studied in \cref{sec:expected_rate}. In \cref{sec:CVaR}, we define the $\beta$-\ac{CVaR} rate performance measure and characterize the sample complexity. In \cref{sec:mirror_descent}, we describe a mirror-descent-based algorithm for empirical $\beta$-\ac{CVaR} rate maximization. Meta-learning is introduced in \cref{sec:meta_learning} as a means to reduce sample complexity. In \cref{sec:numerical}, we present numerical results in order to evaluate the expected rate and $\beta$-\ac{CVaR} rate for layer allocation, and to assess the impact of meta-learning on performance. Finally, in \cref{sec:conclusion}, we conclude the paper and highlight some open problems.

\emph{Notation:} 
Random variables and vectors are denoted by lowercase and boldface lowercase Roman-font letters, respectively. Realizations of random variables and vectors are denoted by lowercase and boldface lowercase italic-font letters, respectively. For example, $x$ is a realization of random variable $\mathrm{x}$ and $\bm{x}$ is a realization of random vector $\mathbf{x}$.
For any positive integer $K$, we define the set $[K]\triangleq \{1,2,\ldots,K\}$. 
The cardinality and convex hull of a set $\mathcal L$ are denoted by $|\mathcal{L}|$ and $\text{conv}(\mathcal L)$, respectively. 
The $\ell^1$-norm and $\ell^2$-norm of a vector $\bm{s}$ are denoted by $\norms{\bm{s}}_1$ and $\norms{\bm{s}}_2$, respectively.
For two scalars $a$ and $b$, the indicator of the event $a\geq b$ is denoted by $\bm{1}_{a\geq b}$. That is, $\bm{1}_{a\geq b}$ equals one if $a\geq b$ and zero otherwise.
The maximum between real scalar $r$ and zero is denoted by $(r)^+$. 
The set of non-negative real numbers is denoted by $\mathbb R_+$. The nearest positive integer to scalar $x$ is denoted by $\nint{x}$. 
$\diag(\bm{u})$ represents a diagonal matrix with diagonal given by the vector $\bm{u}$.

\section{System Model and Problem Definition}\label{sec:model}
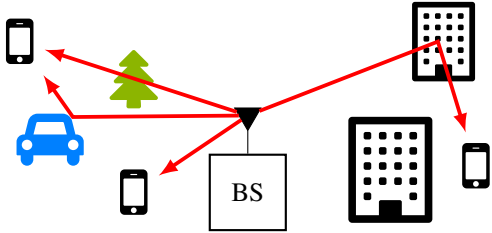
\begin{figure}[!t]
	\centering
	\begin{tikzpicture}[>=latex,
	ant/.style={%
		draw,
		fill,
		regular polygon,
		regular polygon sides=3,
		shape border rotate=180,
	},
]

\node (BS) [thick,draw,minimum width=1cm,minimum height=1cm, font=\small] at (0,0) {BS};
\node (Ant1) [ant, above=0.3cm of BS,scale=0.5] {};
\path[draw] (BS.north) -- (Ant1);

\node (building1)  at (1.8,0.3) {{\fontsize{40}{50} \faBuildingO}};
\node (building2)  at (2.5,2) {{\fontsize{30}{30} \faBuildingO}};

\node (client1)  at (3.0,0.35){{\Huge \faMobile}};
\node (client2)  at (-3,2){{\Huge \faMobile}};
\node (client3)  at (-1.5,0){{\Huge \faMobile}};

\definecolor{applegreen}{rgb}{0.55, 0.71, 0.0}
\node[applegreen] (tree)  at (-1.5,1.5) {{\huge \faTree}};

\definecolor{azure(colorwheel)}{rgb}{0.0, 0.5, 1.0}
\node[azure(colorwheel)] (car)  at (-2.6,0.7) {{\huge \faCar}};

\definecolor{bostonuniversityred}{rgb}{0.8, 0.0, 0.0}
\coordinate (A) at ($(Ant1)+(2mm,0)$);
\path[draw,->,line width=0.5mm,red] (Ant1) -- (building2.center) -- (client1);
\path[draw,->,line width=0.5mm,red] (Ant1) -- (client2);
\path[draw,->,line width=0.5mm,red] (Ant1) -- ($(car.center)+(0.3,0.3)$) -- (client2);
\path[draw,->,line width=0.5mm,red] (Ant1) -- (client3);

\end{tikzpicture}
	\caption{Illustration of the broadcast setting under study. A single-antenna base station (BS) broadcasts a common message to single-antenna clients. The signal to each client undergoes a fading channel in which the fading coefficient is drawn from a common fading distribution $p_{\mathrm{h}}(h)$. This paper is concerned with the average rate of the $\beta$-fraction of users with the worst instantaneous channel conditions (see \cref{fig:cvar}).}
	\label{fig:model}
\end{figure}

We consider the system depicted in \cref{fig:model} in which a single-antenna \ac{BS} broadcasts a common message to single-antenna clients over a fading broadcast channel. The fading coefficient for each client is drawn from a common fading distribution $p_{\mathrm{h}}(h)$, and is assumed to remain constant for the duration of a coding block consisting of $n$ symbols. The common fading distribution $p_{\mathrm{h}}(h)$ may take the form of a mixture model, as in \cite{ntranos2009multicast}, in order to account for heterogeneous long-term effects such as path loss and shadowing. 

The signal received by a client at time $t\in[n]$, denoted by $\mathrm{y}(t)$, can be expressed as
\begin{IEEEeqnarray}{c}\label{eq:bc}
	\mathrm{y}(t)=\sqrt{P}\mathrm{h} \mathrm{x}(t)+\mathrm{z}(t),
\end{IEEEeqnarray}
where $P>0$ denotes the \ac{BS} transmission power; $\mathrm{x}(t)\in\mathbb C$ denotes the signal transmitted at time $t$, which is subject to the average power constraint
\begin{IEEEeqnarray}{c}\label{eq:power_constraint}
	\mathbb E\sqb{|\mathrm{x}(t)|^2}\leq 1;
\end{IEEEeqnarray}
channel coefficient $\mathrm{h}\sim p_{\mathrm{h}}(h)$ denotes the quasi-static fading coefficient; and $\mathrm{z}(t)\sim\mathcal{CN}(0,1)$ denotes the \ac{AWGN}. 

We assume that the \ac{BS} does not know the fading realizations nor the common fading distribution $p_{\mathrm{h}}(h)$, while each client knows its own channel $\mathrm{h}$.
Due to the lack of \ac{CSI}, the \ac{BS} applies \acf{LDM} \cite{shamai2003broadcast} with $M$ layers, or sub-messages, in order to enable differential quality of service at the clients.  The transmitted signal $\mathrm{x}(t)$ in \eqref{eq:bc} is accordingly given as
\begin{IEEEeqnarray}{c}
	\mathrm{x}(t)=\sum_{m=1}^{M}\mathrm{x}_m(t),
\end{IEEEeqnarray}
where $\mathrm{x}_m(t)\sim\mathcal{CN}(0,\lambda_m)$, with $m\in[M]$, denotes a symbol from a Gaussian random codebook with average power $\lambda_m$ that is used to encode sub-message $\mathrm{w}_m\in[2^{n\rho_m}]$ of rate $\rho_m\geq 0$. To satisfy the normalized power constraint in \eqref{eq:power_constraint}, the power-allocation vector $\pmb{\lambda}\triangleq (\lambda_1,\ldots,\lambda_M)$ must thus lie in the simplex 
\begin{IEEEeqnarray}{c}\label{eq:def_simplex}
	\Delta_c^M\triangleq \cb{\pmb{\lambda}\in\mathbb R_+^M:\sum_{m=1}^M\lambda_m\leq1}.
\end{IEEEeqnarray}
We refer to message $\mathrm{w}_m$ and corresponding encoded signal $\mathrm{x}_m(t)$ as the $m$th layer.

\looseness=-1
Each client decodes sub-messages by applying \ac{SCD} with the order $\mathrm{w}_1,\ldots,\mathrm{w}_M$. When decoding layer $m\in[M]$, all subsequent layers are treated as \ac{AWGN}. Each client can hence decode only a subset of layers depending on its channel gain $\mathrm{g}\triangleq |\mathrm{h}|^2$. We denote by $I_m\triangleq \sum_{i=m+1}^{M}\lambda_i$ the normalized power level of the inter-layer interference affecting the decoding of layer $m$, and as $p_{\mathrm g}(g)$ the distribution of the channel gain $\mathrm{g}$.

We parametrize the rate $\rho_m$ of layer $m$ as \cite{shamai2003broadcast}
\begin{IEEEeqnarray}{c}\label{eq:rate_layer}
	\rho_m(\bm{s}^m,\pmb{\lambda})\triangleq \log_2\rb{1+\frac{\norms{\bm{s}^m}_1\lambda_m P}{1+ \norms{\bm{s}^m}_1I_m P}},
\end{IEEEeqnarray}
\looseness=-1
where $\bm{s}\triangleq(s_1,\ldots,s_M)\in\mathbb R_+^M$ is a non-negative vector set by the \ac{BS}, and vector $\bm{s}^m\triangleq (s_1,\ldots,s_m)\in\mathbb R_+^{m}$ consists of the first $m$ elements of $\bm{s}$. Assuming that all previous layers are correctly decoded, the rate achievable for layer $m$ by a client with channel gain $\mathrm{g}$ is $\log_2(1+\mathrm{g}\lambda_m P/(1+\mathrm{g}I_m P))$. Therefore, the client can decode all layers up to layer $m$ if and only if its channel gain satisfies the inequality $\mathrm{g}\geq\norms{\bm{s}^m}_1$.
Accordingly, given the power and rate allocation vectors $\pmb{\lambda}$ and $\bm{s}$, the total rate that can be decoded by a client with channel gain $g$ is given as
\begin{IEEEeqnarray}{c}\label{eq:def_total_rate}
	R(\bm{s},\pmb{\lambda},g)\triangleq \sum_{m=1}^{M}\rho_m(\bm{s}^m,\pmb{\lambda})\bm{1}_{g\geq \norms{\bm{s}^m}_1}.
\end{IEEEeqnarray}

\looseness=-1
We study the optimization of the rate and power allocation vectors $(\bm{s},\pmb{\lambda})$ under two performance metrics. Specifically, we first consider the optimization of the \emph{expected achievable rate} 
\begin{IEEEeqnarray}{c}\label{eq:cvar_expected_rate}
	\bar{R}(\bm{s},\pmb{\lambda})\triangleq\mathbb E_{\mathrm g}\sqb{R(\bm{s},\pmb{\lambda},\mathrm{g})},
\end{IEEEeqnarray}
where the expectation is over the fading distribution $p_{\mathrm{g}}(g)$, in \cref{sec:expected_rate}. Then, in \cref{sec:CVaR}, we investigate a more general metric, the \ac{CVaR}, which can naturally account for the performance of ultra-reliable communication.

\section{Expected Achievable Rate}\label{sec:expected_rate}
In this section, we study the maximization of the expected achievable rate $\bar{R}(\bm{s},\pmb{\lambda})$ in \eqref{eq:cvar_expected_rate} over the power and rate allocation vectors $\pmb{\lambda}$ and $\bm{s}$. That is, we consider the optimization problem
\begin{IEEEeqnarray}{c}\label{eq:opt_prob_1}
	(\bm{s}^*,\pmb{\lambda}^*)\in\argmax_{(\bm{s},\pmb{\lambda})\in \mathbb R_+^M\times\Delta_c^M}\bar{R}(\bm{s},\pmb{\lambda}).
\end{IEEEeqnarray}
The optimization problem \eqref{eq:opt_prob_1} depends on the unknown fading distribution $p_{\mathrm{g}}(g)$. 
In this section, we first review known optimality results for the case in which $p_{\mathrm{g}}(g)$ is known, and then we address the setting of interest in which the distribution $p_{\mathrm{g}}(g)$ in unavailable.

\subsection{Known Channel Distribution}
In \cite{shamai2003broadcast}, problem \eqref{eq:opt_prob_1} was studied under the assumption that the \ac{BS} knows the fading distribution $p_{\mathrm{g}}(g)$, and the optimal power allocation density was derived for an infinite number of layers $M\to\infty$. For reference, we report the main result in the following proposition.
\begin{proposition}[\!\!{\cite[Sec. II.B]{shamai2003broadcast}}]\label{prop:shamai}
	The expected rate achieved when the \ac{BS} knows the fading distribution $p_{\mathrm{g}}(g)$ and \ac{LDM} is applied with an infinite number of layers ($M\to\infty$) is given as
	\begin{IEEEeqnarray}{c}
		\bar{R}_\infty=\int_{0}^{\infty}\Pr\sqb{\mathrm{g}\geq u}\frac{u\rho(u)}{1+uI(u)}\diff{u},
	\end{IEEEeqnarray}
	with power allocation density $\rho(u)=-\diff I(u)/\diff u$ and accumulated interference
	\begin{IEEEeqnarray}{c}\label{eq:shamai_I}
		I(u)=\left\lbrace\begin{array}{ll}
			\frac{\Pr\sqb{\mathrm{g}\geq u}-up_{\mathrm{g}}(u)}{u^2p_{\mathrm{g}}(u)},&\text{for }u_0\leq u\leq u_1,\\
			0,&\text{otherwise},
		\end{array}\right.
	\end{IEEEeqnarray}
	where $u_0$ is determined by the equality $I(u_0)=P$, and $u_1$ is determined by the equality $I(u_1)=0$.
\end{proposition}

Note that the result \eqref{eq:shamai_I} holds only for continuous fading distributions $p_{\mathrm{g}}(g)$.
For the special case of Rayleigh fading, this result can be specialized as follows. 
\begin{corollary}[\!\!{\cite[Sec. II.C]{shamai2003broadcast}}]\label{cor:shamai}
	For Rayleigh fading, i.e., for $\mathrm{h}\sim\mathcal{CN}(0,1)$, the expected rate achieved when the \ac{BS} knows the fading distribution $p_{\mathrm{g}}(g)$ and \ac{LDM} is applied with an infinite number of layers ($M\to\infty$) can be expressed as
	\begin{IEEEeqnarray}{c}\label{eq:expected_rate_shamai}
		\bar{R}_\infty=2\int_{u_0}^{\infty}\frac{\exp(-u)}{u}\diff{u}-2\int_{1}^{\infty}\frac{\exp(-u)}{u}\diff{u}-\rb{\exp(-u_0)-\exp(-1)},
	\end{IEEEeqnarray}
	where
	\begin{IEEEeqnarray}{c}
		u_0=\frac{2}{1+\sqrt{1+4P}}.
	\end{IEEEeqnarray}
\end{corollary}

To the best of our knowledge, finding an explicit solution to problem \eqref{eq:opt_prob_1} for a finite number of layers is an open problem even when the fading distribution $p_{\mathrm{g}}(g)$ is known. That said, the expected achievable rate in \cref{prop:shamai} can be viewed as an upper bound on the expected rate achieved for a finite number of layers and for unknown fading distribution. 

\subsection{Unknown Channel Distribution: Empirical Average Rate Maximization}
In this paper, we assume that the \ac{BS} does not know the fading distribution $p_{\mathrm{g}}(g)$, and hence it cannot directly optimize the expected achievable rate $\bar{R}(\bm{s},\pmb{\lambda})$. Instead, we assume that the \ac{BS} has access to a dataset
\begin{IEEEeqnarray}{c}\label{eq:training_set}
	\mathcal G =\{g_1,\ldots,g_N\}
\end{IEEEeqnarray}
consisting of $N$ fading realizations sampled in an \ac{iid} manner from distribution $p_{\mathrm{g}}(g)$. 
Based on dataset $\mathcal G$, the \ac{BS} approximates the expected achievable rate with the empirical average
\begin{IEEEeqnarray}{rCl}\label{eq:empir_exp_rate}
	\bar{R}^{\mathcal{G}}(\bm{s},\pmb{\lambda})&=&\frac{1}{N}\sum_{i=1}^{N}R(\bm{s},\pmb{\lambda},g_i).
\end{IEEEeqnarray}
The maximization of the average rate \eqref{eq:empir_exp_rate} over power and rate allocation vectors $\pmb{\lambda}$ and $\bm{s}$ can be expressed as the optimization problem
\begin{IEEEeqnarray}{c}\label{eq:opt_prob_empir_1}
	(\bm{s}^{\mathcal G},\pmb{\lambda}^{\mathcal G})\in\argmax_{(\bm{s},\pmb{\lambda})\in \mathbb R_+^M\times\Delta_c^M}\bar{R}^{\mathcal G}(\bm{s},\pmb{\lambda}).
\end{IEEEeqnarray}
A solution to problem \eqref{eq:opt_prob_empir_1} can be practically obtained via an iterative optimization scheme as detailed in \cref{sec:mirror_descent}.

We emphasize that optimizing the average rate $\bar{R}^{\mathcal{G}}(\bm{s},\pmb{\lambda})$ via problem \eqref{eq:opt_prob_empir_1} is useful not only when the fading distribution $p_{\mathrm{g}}(g)$ is unknown, but also when the direct optimizations in \eqref{eq:opt_prob_1} based on knowledge of the distribution $p_{\mathrm g}(g)$ is not tractable. In this latter case, one can potentially generate the dataset $\mathcal G$ with an arbitrary number of fading realizations $N$.

\subsection{Optimality Gap and Sample Complexity}\label{sec:expected_rate_sample_complexity}
An important theoretical question is whether the expected achievable rate obtained under the power and rate allocation vectors \eqref{eq:opt_prob_empir_1} approaches the ground-truth maximum expected achievable rate obtained with vectors \eqref{eq:opt_prob_1} as the size of the dataset increases. If so, it would also be interesting to quantify how many samples $N$ are required to achieve a desired level of approximation. This is the subject of this subsection.

To proceed,  we define the optimality gap 
\begin{IEEEeqnarray}{c}\label{eq:gen_error}
	e^{\mathcal G}\triangleq \bar{R}\big(\bm{s}^*,\pmb{\lambda}^*\big)-\bar{R}\big(\bm{s}^{\mathcal G},\pmb{\lambda}^{\mathcal G}\big)
\end{IEEEeqnarray}
as the difference between the expected rate achieved with optimal power and rate allocation vectors \eqref{eq:opt_prob_1} and the expected rate achieved by the empirical rate maximization \eqref{eq:opt_prob_empir_1}. The optimality gap is random due to the stochastic nature of the dataset $\mathcal G$. 

To bound the optimality gap, we assume that the norms of the optimal vectors $\bm{s}^*$ and $\bm{s}^{\mathcal G}$ in \eqref{eq:opt_prob_1} and \eqref{eq:opt_prob_empir_1}, respectively, can be bounded as $\max\{\norms{\bm{s}^*}_1,\norms{\bm{s}^{\mathcal G}}_1\}\leq S$ for some known constant $S>0$. Note that this assumption is not restrictive since, in practice, $S$ represents the largest fading gain $g$ that a client is expected to experience. The following proposition bounds the optimality gap under this assumption. 

\begin{proposition}\label{prop:expected_rate_convergence}
	Let $\mathcal G=\{g_1,\ldots,g_N\}$ be a dataset of $N$ fading realizations drawn independently from the fading distribution $p_{\mathrm{g}}(g)$, and let $\delta\in(0,1]$. With probability at least $1-\delta$, the optimality gap \eqref{eq:gen_error} is bounded, for rate allocation vectors with bounded norms $\max\{\norms{\bm{s}^*}_1,\norms{\bm{s}^{\mathcal G}}_1\}\leq S$, as
	\begin{IEEEeqnarray}{c}\label{eq:bound_opt_gap_1}
		e^{\mathcal G}\leq  \rb{4\sqrt{\frac{(2N+1)\ln(N+1)}{3N(N+1)}}+\sqrt{\frac{2\ln(2/\delta)}{N}}}2\log_2\rb{1+SP}.
	\end{IEEEeqnarray}
\end{proposition}
\begin{IEEEproof}
	See Appendix \ref{app:proof_prop_expected_rate_convergence}.
\end{IEEEproof}

This result shows that the optimality gap scales with number of data points, $N$, as $\mathcal{O}(\sqrt{\ln(N)/N})$, implying that any level of accuracy can be attained as the dataset grows larger, i.e., as $N\rightarrow\infty$. Furthermore, for a given desired optimality gap $e^{\mathcal G}\leq\epsilon$, the required number of data points $N$, i.e., the sample complexity, satisfies the approximate inequality
\begin{IEEEeqnarray}{c}
	\frac{N}{\ln(N)}\gtrapprox\rb{\frac{\log_2(SP)}{\epsilon}}^2 
\end{IEEEeqnarray}
for large $N$. Intuitively, the sample complexity increases with the \ac{SNR} metric $SP$ since, as the achievable rate increases, a better approximation is required to achieve the same subtractive optimality gap.

\section{Conditional Value at Risk}\label{sec:CVaR}
In this section, we move beyond the expected rate metric with the goal of investigating a performance measure that is closer to the requirements of ultra-reliable systems, namely the \acf{CVaR} rate.

\subsection{Outage Rate and \ac{CVaR} Rate}
The standard performance measure used for ultra-reliable systems is the outage rate \cite{angjelichinoski2019statistical,jurdi2018variable,popovski2019wireless,hampel2019ultra}. For outage probability $\beta\in[0,1]$ and vectors $\bm{s}$ and $\pmb{\lambda}$, the $\beta$-outage rate $r_\beta(\bm{s},\pmb{\lambda})$ is the largest total rate in \eqref{eq:def_total_rate} that can be guaranteed with probability $1-\beta$. Mathematically, it is defined as
\begin{IEEEeqnarray}{c}\label{eq:def_var}
	r_\beta(\bm{s},\pmb{\lambda})\triangleq \max\cb{r\in\mathbb R_+:\Pr\sqb{R(\bm{s},\pmb{\lambda},\mathrm{g})\geq r}\geq 1-\beta}.
\end{IEEEeqnarray}
For the broadcast setting under study, a $\beta$-outage rate $r_\beta(\bm{s},\pmb{\lambda})=r$ indicates that a fraction $1-\beta$ of all possible clients is guaranteed to attain a rate at least equal to $r$; or, conversely, that a fraction $\beta$ of all possible clients cannot decode at rate $r$, as illustrated in \cref{fig:cvar}.

The $\beta$-outage rate does not provide any information about the rates achieved by the $\beta$-fraction of users with the worst channel gain. To obtain a more refined analysis of the $\beta$-fraction of the least performing users, in this paper, as illustrated in \cref{fig:cvar}, we introduce the \emph{$\beta$-\ac{CVaR} rate} $R_\beta(\bm{s},\pmb{\lambda})$ as the expected rate achieved by the $\beta$-fraction of clients with the worst channels. 
Formally, the $\beta$-\ac{CVaR} rate is defined as 
\begin{IEEEeqnarray}{rCl}\label{eq:def_cvar}
	R_\beta(\bm{s},\pmb{\lambda})&\triangleq& \condExU{\mathrm g}{R(\bm{s},\pmb{\lambda},\mathrm{g})}{R(\bm{s},\pmb{\lambda},\mathrm{g})\leq r_\beta(\bm{s},\pmb{\lambda})},
\end{IEEEeqnarray}
where the expectation is conditioned on the event that the rate is lower than the $\beta$-outage rate.

For $\beta=0$, we have $r_0(\bm{s},\pmb{\lambda})=R_0(\bm{s},\pmb{\lambda})=0$ for all $\bm{s}\in\mathbb R_+^M$ and $\pmb{\lambda}\in\Delta_c^M$, so, we limit the range of $\beta$ to $\beta\in(0,1]$.
Furthermore, at the other extreme, for $\beta=1$, the 1-\ac{CVaR} rate can be seen to coincide with the expected achievable rate, i.e., $R_1(\bm{s},\pmb{\lambda})=\bar{R}(\bm{s},\pmb{\lambda})$.
In practice, for ultra-reliable applications, smaller values of $\beta$ are typically of interest. 

For any $\beta\in(0,1)$, the $\beta$-\ac{CVaR} rate can be expressed via the \emph{variational representation} \cite[Theorem 1]{rockafellar2000optimization}
\begin{IEEEeqnarray}{c}\label{eq:cvar_variational}
	R_\beta(\bm{s},\pmb{\lambda})=\max_{r\in\mathbb R_+}f_\beta(\bm{s},\pmb{\lambda},r),
\end{IEEEeqnarray}
where the function $f_\beta(\bm{s},\pmb{\lambda},r)$ is defined as
\begin{IEEEeqnarray}{c}\label{eq:def_f}
	f_\beta(\bm{s},\pmb{\lambda},r)\triangleq r-\beta^{-1}\mathbb E_{\mathrm g}\sqb{\rb{r-R(\bm{s},\pmb{\lambda},\mathrm{g})}^{+}}.
\end{IEEEeqnarray}
Furthermore, the maximum in \eqref{eq:cvar_variational} is attained at the $\beta$-outage rate $r=r_\beta(\bm{s},\pmb{\lambda})$, i.e.,
\begin{IEEEeqnarray}{c}
	R_\beta(\bm{s},\pmb{\lambda})=f_\beta(\bm{s},\pmb{\lambda},r_\beta(\bm{s},\pmb{\lambda})).
\end{IEEEeqnarray}

Finally, the maximization of the $\beta$-\ac{CVaR} rate over the power and rate allocation vectors $\pmb{\lambda}$ and $\bm{s}$ can be expressed as the joint optimization problem
\begin{IEEEeqnarray}{c}\label{eq:opt_prob_beta}
	(\bm{s}_\beta^*,\pmb{\lambda}_\beta^*)\in\argmax_{(\bm{s},\pmb{\lambda})\in \mathbb R_+^M\times\Delta_c^M}R_\beta(\bm{s},\pmb{\lambda}) = \argmax_{(\bm{s},\pmb{\lambda})\in \mathbb R_+^M\times\Delta_c^M}f_\beta(\bm{s},\pmb{\lambda},r_\beta(\bm{s},\pmb{\lambda})).
\end{IEEEeqnarray}
To the best of our knowledge, finding an explicit solution to the maximization \eqref{eq:opt_prob_beta} for any $\beta\in(0,1)$ is an open problem even when the fading distribution $p_{\mathrm{g}}(g)$ is known. 

\subsection{Empirical \ac{CVaR} Maximization}
Since the \ac{BS} does not know the fading distribution $p_{\mathrm{g}}(g)$, it cannot directly optimize the $\beta$-\ac{CVaR} rate. Instead, similar to the expected rate maximization problem studied in \cref{sec:expected_rate}, the \ac{BS} approximates the $\beta$-\ac{CVaR} rate with an empirical average over the dataset $\mathcal G$ in \eqref{eq:training_set}. Specifically, following the variational representation \eqref{eq:cvar_variational}, the \emph{empirical $\beta$-\ac{CVaR} rate}, for any $\beta\in(0,1)$, is defined as
\begin{IEEEeqnarray}{c}\label{eq:empir_cvar}
	R^{\mathcal G}_{\beta}(\bm{s},\pmb{\lambda})\triangleq \max_{r\in\mathbb R_+}f^{\mathcal G}_\beta(\bm{s},\pmb{\lambda},r),
\end{IEEEeqnarray}
with $f^{\mathcal G}_\beta(\bm{s},\pmb{\lambda},r)$ being the empirical approximation of function $f_\beta(\bm{s},\pmb{\lambda},r)$ in \eqref{eq:def_f}, i.e.,
\begin{IEEEeqnarray}{rCl}\label{eq:ampir_f}
	f^{\mathcal G}_\beta(\bm{s},\pmb{\lambda},r)&=&r-\frac{1}{N\beta}\sum_{i=1}^{N}\rb{r-R(\bm{s},\pmb{\lambda},g_i)}^{+}.
\end{IEEEeqnarray}
Furthermore, we define the \emph{empirical $\beta$-outage rate} $r_\beta^{\mathcal G}(\bm{s},\pmb{\lambda})$ as the optimal $r$ for problem \eqref{eq:empir_cvar}. Hence, we have
\begin{IEEEeqnarray}{c}
	R^{\mathcal G}_{\beta}(\bm{s},\pmb{\lambda})=f^{\mathcal G}_\beta\big(\bm{s},\pmb{\lambda},r_\beta^{\mathcal G}(\bm{s},\pmb{\lambda})\big).
\end{IEEEeqnarray}
Overall, the maximization of the empirical $\beta$-\ac{CVaR} rate over power and rate allocation vectors $\pmb{\lambda}$ and $\bm{s}$ can hence be expressed as the optimization problem
\begin{IEEEeqnarray}{c}\label{eq:opt_prob_empir_beta}
	(\bm{s}_\beta^{\mathcal G},\pmb{\lambda}_\beta^{\mathcal G})\in\argmax_{(\bm{s},\pmb{\lambda})\in \mathbb R_+^M\times\Delta_c^M}R_\beta^{\mathcal G}(\bm{s},\pmb{\lambda}) = \argmax_{(\bm{s},\pmb{\lambda})\in \mathbb R_+^M\times\Delta_c^M}f_\beta^{\mathcal G}\big(\bm{s},\pmb{\lambda},r_\beta^{\mathcal G}(\bm{s},\pmb{\lambda})\big).
\end{IEEEeqnarray}

In closing this section, we observe that, unlike the $\beta$-outage rate \eqref{eq:def_var}, the empirical $\beta$-outage rate $r_\beta^{\mathcal G}(\bm{s},\pmb{\lambda})$ has a closed-form expression. Defining as $g_{[i]}$ the $i$th smallest channel gain in dataset $\mathcal{G}$, i.e., $g_{[1]}\leq g_{[2]}\leq\cdots\leq g_{[N]}$, we have the following proposition. 
\begin{proposition}\label{prop:opt_r}
	For any  power and rate allocation vectors $\pmb{\lambda}\in\Delta_c^M$ and $\bm{s}\in\mathbb R_+^M$, the empirical $\beta$-outage rate is given as $r_\beta^{\mathcal G}(\bm{s},\pmb{\lambda})= R(\bm{s},\pmb{\lambda},g_{[\nint{N\beta}]})$.
\end{proposition}
\begin{IEEEproof}
	See Appendix \ref{app:proof_opt_r}.
\end{IEEEproof}

By \cref{prop:opt_r}, the objective in \eqref{eq:opt_prob_empir_beta} can be expressed as
\begin{IEEEeqnarray}{c}\label{eq:explicit_empirical_cvar}
	R_\beta^{\mathcal G}(\bm{s},\pmb{\lambda})=\frac{\nint{N\beta}}{N\beta}\bar{R}^{\mathcal G_\beta}(\bm{s},\pmb{\lambda})+\rb{1-\frac{\nint{N\beta}}{N\beta}}R(\bm{s},\pmb{\lambda},g_{[\nint{N\beta}]}),
\end{IEEEeqnarray}
where we have defined the subset
\begin{IEEEeqnarray}{c}
	\mathcal G_\beta\triangleq \{g_{[1]},g_{[2]},\ldots,g_{[\nint{N\beta}]}\}\subseteq \mathcal G,
\end{IEEEeqnarray}
which consists of the lowest $\beta$-fraction of channel gains.
Therefore, the $\beta$-\ac{CVaR} rate can be obtained from the expected achievable rate calculated for the subset $\mathcal G_\beta$, with a minor correction (the second term in \eqref{eq:explicit_empirical_cvar}) if the product $N\beta$ is not an integer.

\subsection{Optimality Gap and Sample Complexity}\label{sec:CVaR_sample_complexity}
In this section, we study the sample complexity for the $\beta$-\ac{CVaR} rate metric. To this end, for any $\beta\in (0,1)$, we define the optimality gap 
\begin{IEEEeqnarray}{c}\label{eq:gen_error_beta}
	e_\beta^{\mathcal G}\triangleq R_{\beta}\big(\bm{s}_\beta^*,\pmb{\lambda}_\beta^*\big)-R_{\beta}\big(\bm{s}_\beta^{\mathcal G},\pmb{\lambda}_\beta^{\mathcal G}\big)
\end{IEEEeqnarray}
as the difference between the $\beta$-\ac{CVaR} rate achieved with optimal power and rate allocation vectors \eqref{eq:opt_prob_beta} and the $\beta$-\ac{CVaR} rate achieved by the empirical maximization \eqref{eq:opt_prob_empir_beta}.

Similar to the case of expected rate maximization studied in \cref{sec:expected_rate_sample_complexity}, we assume rate allocation vectors with bounded norms, i.e., $\max\{\norms{\bm{s}_\beta^*}_1,\norms{\bm{s}_\beta^{\mathcal G}}_1\}\leq S$. The following proposition bounds optimality gap \eqref{eq:gen_error_beta} under this assumption.
\begin{proposition}\label{prop:f_convergence}
	Let $\mathcal G=\{g_1,\ldots,g_N\}$ be a dataset of $N$ fading realizations drawn independently from the fading distribution $p_{\mathrm{g}}(g)$, and let $\delta\in(0,1]$. With probability at least $1-\delta$, the optimality gap \eqref{eq:gen_error_beta} is bounded, for $\beta\in(0,1)$ and rate allocation vectors with bounded norms $\max\{\norms{\bm{s}_\beta^*}_1,\norms{\bm{s}_\beta^{\mathcal G}}_1\}\leq S$, as
	\begin{IEEEeqnarray}{c}\label{eq:bound_opt_gap_beta}
		e_\beta^{\mathcal G}\leq \beta^{-1}\Bigg(4\sqrt{\frac{(2N+1)\ln(N+1)}{3N(N+1)}}+\sqrt{\frac{2\ln(2/\delta)}{N}}\Bigg)2\log_2(1+SP).
	\end{IEEEeqnarray}
\end{proposition}
\begin{IEEEproof}
	See Appendix \ref{app:proof_prop_f_conv}.
\end{IEEEproof}

For any fixed $\beta\in(0,1)$, \cref{prop:f_convergence} demonstrates that, in a manner similar to \cref{prop:expected_rate_convergence}, any level of accuracy can be attained as $N\to\infty$. Furthermore, the  scaling of the optimality gap in terms of number of data points, $N$, and \ac{SNR} metric, $SP$, is equivalent to that described for expected rate maximization in \cref{prop:expected_rate_convergence}. However, \cref{prop:f_convergence} also shows that the number of samples required to maintain a desired optimality gap $\epsilon$ increases as $\beta$ decreases according to the approximate inequality
\begin{IEEEeqnarray}{c}\label{eq:sample_complexity_beta}
	\frac{N}{\ln(N)}\gtrapprox\rb{\frac{\log_2(SP)}{\beta\epsilon}}^2
\end{IEEEeqnarray}
for sufficiently large $N$. Intuitively, this is because, as $\beta$ becomes smaller, the $\beta$-\ac{CVaR} rate is calculated with respect to increasingly rarer outage events, which require more data to be observed in sufficient numbers.

\section{Mirror Gradient Descent for $\beta$-\ac{CVaR} Rate Maximization}\label{sec:mirror_descent}
In this section, we introduce a gradient-based iterative optimization procedure to tackle the empirical $\beta$-\ac{CVaR} rate maximization problem \eqref{eq:opt_prob_empir_beta}. The approach is based on the introduction of a surrogate smooth objective and on mirror descent, as described in the rest of this section and summarized in \cref{alg:empirical_optimization_beta}.
\begin{algorithm}
	\DontPrintSemicolon
	\caption{Empirical \ac{CVaR} maximization}
	\label{alg:empirical_optimization_beta}
	\SetKwInOut{Input}{Input}
	\Input{Dataset $\mathcal{G}$, fraction $\beta\in(0,1]$}
	\KwIn{Initialize $\bm{u}\in\mathbb R^M$ and $\pmb{\lambda}\in\Delta_c^M$}
	set $i=0$ \;
	set $\bm{u}^{(i)}=\bm{u}$ and $\pmb{\lambda}^{(i)}=\pmb{\lambda}$ \;
	\While{not converged}
	{
		set $i\gets i+1$ \;
		set $\bm{u}^{(i)}\gets \text{GD}_\beta(\bm{u}^{(i-1)};\mathcal{G},\pmb{\lambda}^{(i-1)})$  (defined in \eqref{eq:gradient_descent})\;
		set $\pmb{\lambda}^{(i)} \gets \text{EG}_\beta(\pmb{\lambda}^{(i-1)};\mathcal{G},\bm{u}^{(i-1)})$  (defined in \eqref{eq:exponentiated_gradient}) \;
	}
	\Return $\big(\bm{s}^{(i)}=\exp(\bm{u}^{(i)}),\pmb{\lambda}^{(i)}\big)$
\end{algorithm}

\subsection{Smooth Surrogate Objective}\label{sec:base_learner}
A first challenge in developing iterative solutions to problem \eqref{eq:opt_prob_empir_beta}, is that the partial derivative of the indicator in the achievable rate expression \eqref{eq:def_total_rate} with respect to vector $\bm{s}$ equals zero almost everywhere. Therefore, in order to facilitate the application of a gradient-based optimization procedure, we replace the rate $R(\bm{s},\pmb{\lambda},g)$ in \eqref{eq:def_total_rate} with the smooth surrogate objective
\begin{IEEEeqnarray}{c}\label{eq:sigmoid_based_surrogate}
	R_\sigma(\bm{s},\pmb{\lambda},g) \triangleq \sum_{m=1}^{M}\rho_m(\bm{s}^m,\pmb{\lambda})\sigma(c(g-\norms{\bm{s}^m}_1)),
\end{IEEEeqnarray}
where $\sigma(x)\triangleq 1/(1+\exp(-x))$ is the sigmoid function, and the parameter $c>0$ determines the trade-off between smoothness and accuracy of the surrogate approximation. As $c\to\infty$, the surrogate \eqref{eq:sigmoid_based_surrogate} tends uniformly to the original rate \eqref{eq:def_total_rate}, while smaller values of $c$ yield non-zero partial derivatives with respect to $\bm{s}$.

Using the approximation \eqref{eq:sigmoid_based_surrogate}, we define the \emph{surrogate empirical \ac{CVaR} maximization problem} as
\begin{IEEEeqnarray}{c}\label{eq:opt_surrogate_beta}
	\big(\tilde{\bm{s}}_\beta^{\mathcal G},\tilde{\pmb{\lambda}}_\beta^{\mathcal G}\big)=\argmax_{(\bm{s},\pmb{\lambda})\in \mathbb R_+^M\times\Delta_c^M}\tilde{R}_\beta^{\mathcal G}(\bm{s},\pmb{\lambda}),
\end{IEEEeqnarray}
where, based on \eqref{eq:explicit_empirical_cvar}, we have defined
\begin{IEEEeqnarray}{c}\label{eq:explicit_surrogate_empirical_cvar}
	\tilde{R}_\beta^{\mathcal G}(\bm{s},\pmb{\lambda})\triangleq\frac{\nint{N\beta}}{N\beta}\tilde{R}^{\mathcal G_\beta}(\bm{s},\pmb{\lambda})+\rb{1-\frac{\nint{N\beta}}{N\beta}}R_\sigma(\bm{s},\pmb{\lambda},g_{[\nint{N\beta}]})
\end{IEEEeqnarray}
with the \emph{surrogate average rate}
\begin{IEEEeqnarray}{c}\label{eq:surrogate_obj_1}
	\tilde{R}^{\mathcal G_\beta}(\bm{s},\pmb{\lambda})\triangleq \frac{1}{\nint{N\beta}}\sum_{i=1}^{\nint{N\beta}}R_\sigma(\bm{s},\pmb{\lambda},g_{[i]}).
\end{IEEEeqnarray}

\subsection{Mirror Descent}
Although the objective in \eqref{eq:opt_surrogate_beta} is smooth, plain-vanilla gradient descent cannot be applied to address the optimization \eqref{eq:opt_surrogate_beta} due to the domain constraints on the optimization variables $(\bm{s},\pmb{\lambda})\in \mathbb R_+^M\times\Delta_c^M$. To tackle the constraint $\bm{s}\in \mathbb R_+^M$, we parametrize the rate-allocation vector $\bm{s}$ with a vector $\bm{u}\in\mathbb R^M$ as
\begin{IEEEeqnarray}{c}
	\bm{s}=\exp(\bm{u})\triangleq (\exp(u_1),\ldots,\exp(u_M)).
\end{IEEEeqnarray}
Furthermore, to satisfy the constraint $\pmb{\lambda}\in\Delta_c^M$, we consider a mirror-decent based scheme which adapts the updates to the geometry of the simplex $\Delta_c^M$ via the exponentiated gradient \cite{kivinen1997exponentiated}. Overall, this leads to the updates
\begin{IEEEeqnarray}{c}\label{eq:gradient_descent}
	\bm{u}\gets \bm{u}+\eta\diag(\exp(\bm{u}))\left.\nabla_{\bm{s}}\tilde{R}_\beta^{\mathcal G}(\bm{s},\pmb{\lambda})\right\rvert_{\bm{s}=\exp(\bm{u})}\triangleq \text{GD}_\beta(\bm{u};\mathcal{G},\pmb{\lambda})
\end{IEEEeqnarray}
and
\begin{IEEEeqnarray}{c}\label{eq:exponentiated_gradient}
	\lambda_m\gets \frac{\lambda_m\exp\rb{\gamma[\nabla_{\pmb{\lambda}} \tilde{R}_\beta^{\mathcal G}(\exp(\bm{u}),\pmb{\lambda})]_m}}{\sum_{m'=1}^{M}\lambda_{m'}\exp\rb{\gamma[\nabla_{\pmb{\lambda}} \tilde{R}_\beta^{\mathcal G}(\exp(\bm{u}),\pmb{\lambda})]_{m'}}}\triangleq \text{EG}_\beta(\pmb{\lambda};\mathcal{G},\bm{u}),~\forall m\in[M].
\end{IEEEeqnarray}
The resulting procedure to optimize the empirical $\beta$-\ac{CVaR} rate is summarized in \cref{alg:empirical_optimization_beta}. By \eqref{eq:explicit_surrogate_empirical_cvar}, the algorithm can be also applied to maximize the average rate by setting $\beta=1$.

\section{Reducing Sample Complexity Via Meta-Learning}\label{sec:meta_learning}
In \cref{sec:CVaR}, we have shown that, given the focus of the $\beta$-\ac{CVaR} metric on the performance of a small fraction $\beta$ of the clients with the worst channel gains (see \cref{fig:cvar}), the number of samples required to maintain a desired optimality gap increases as $\beta$ decreases (see \eqref{eq:sample_complexity_beta}).
In this section, we propose meta-learning as a means to reduce sample complexity by leveraging historical data from other deployments, each generally characterized by different fading distributions, as detailed next.

\subsection{Setting}
Let $\mathrm{y}_\tau(t)$ be the signal received by a client in a previous deployment described by variable $\tau$. Similar to the model \eqref{eq:bc}, the received signal can be expressed as
\begin{IEEEeqnarray}{c}\label{eq:meta_learn_model}
	\mathrm{y}_\tau(t)=\sqrt{P}\mathrm{h}_\tau\mathrm{x}_\tau(t)+\mathrm{z}_\tau(t),
\end{IEEEeqnarray}
where $\mathrm{x}_\tau(t)$ denotes the signal transmitted by \ac{BS} $\tau$ at time $t\in[n]$; $\mathrm{z}_\tau(t)\sim\mathcal{CN}(0,1)$ denotes the \ac{AWGN}; and $\mathrm{h}_\tau\sim p_{\mathrm{h}_\tau}$ denotes the quasi-static fading coefficient. The dependence of the fading distribution on the deployment variable $\tau$ indicates that different deployments may have distinct channel statistics. As an example, each deployment may be characterized by Rician fading coefficients $\mathrm{h}_{\tau}\sim\mathcal{CN}(\mu,\sigma^2)$ with parameters $\tau=(\mu,\sigma^2)$.

To support meta-learning, we assume that we have access to data from a subset $\mathcal T=\{\tau_1,\ldots,\tau_D\}$ of $D$ deployments. For each deployment $\tau\in\mathcal T$, we specifically have a dataset $\mathcal{G}^\tau\triangleq \{g_{\tau,1},\ldots,g_{\tau,N}\}$ of $N$ fading realizations sampled in an \ac{iid} manner from distribution $p_{\mathrm{g}_\tau}(g_\tau)$, where $\mathrm{g}_\tau\triangleq|\mathrm{h}_\tau|^2$ is the gain coefficient. We refer to the collection of datasets $\mathcal G^{\mathcal T}\triangleq \{\mathcal G^\tau\}_{\tau\in\mathcal T}$ as the \emph{meta-training data}. The goal of meta-learning is to use this meta-training data prior to deploying a new system in order to reduce the amount of data needed to optimize effectively the power and rate allocation vectors for the latter.

\subsection{MAML}
To demonstrate the gain of meta-learning in reducing the sample complexity, we focus on the \ac{MAML} approach introduced in \cite{finn2017model}. In this approach, the meta-training data $\mathcal G^{\mathcal T}$ is used to optimize an initialization $(\bm{s},\pmb{\lambda}) $ for a gradient-based scheme that addresses the $\beta$-\ac{CVaR} optimization \eqref{eq:opt_surrogate_beta}. Specifically, we adapt \ac{MAML} for the gradient-based scheme described in the previous section. 

First, given a random initialization $(\bm{s},\pmb{\lambda})$, with $\bm{s}=\exp(\bm{u})$ for some $\bm{u}\in\mathbb R^M$, the model parameters are individually adapted to each deployment by applying a single update of the gradient-based scheme in \cref{alg:empirical_optimization_beta}. That is, the updated model parameters $(\bm{u}_\tau,\pmb{\lambda}_\tau)$, for deployment $\tau\in\mathcal T$, are given as
\begin{IEEEeqnarray}{c}
	\bm{u}_{\tau}(\bm{u},\pmb{\lambda})= \text{GD}_\beta(\bm{u};\mathcal{G}^\tau,\pmb{\lambda})
\end{IEEEeqnarray}
and
\begin{IEEEeqnarray}{c}
	\pmb{\lambda}_{\tau}(\bm{u},\pmb{\lambda}) = \text{EG}_\beta(\pmb{\lambda};\mathcal{G}^\tau,\bm{u}),
\end{IEEEeqnarray}
where functions $\text{GD}_\beta(\cdot;\cdot,\cdot)$ and $\text{EG}_\beta(\cdot;\cdot,\cdot)$ are defined in \eqref{eq:gradient_descent} and \eqref{eq:exponentiated_gradient}, respectively. Then, the initialization $(\bm{u},\pmb{\lambda})$ is optimized by maximizing the surrogate average empirical $\beta$-\ac{CVaR} across all deployments as
\begin{IEEEeqnarray}{c}\label{eq:opt_meta}
	\max_{(\bm{u},\pmb{\lambda})\in\mathbb R^M\times\Delta_c^M}\phi(\bm{u},\pmb{\lambda}),
\end{IEEEeqnarray} 
where we have defined the function
\begin{IEEEeqnarray}{c}
	\phi(\bm{u},\pmb{\lambda})\triangleq \sum_{\tau\in\mathcal T}\tilde{R}_\beta^{\mathcal G^\tau}(\exp(\bm{u}_\tau(\bm{u},\pmb{\lambda})),\pmb{\lambda}_\tau(\bm{u},\pmb{\lambda})).
\end{IEEEeqnarray}

To address the optimization \eqref{eq:opt_meta}, a gradient-based scheme is applied in which, similar to the scheme described in the previous section, the vector $\bm{u}$ is updated via gradient descent, whereas the power-allocation vector $\pmb{\lambda}$ is updated via mirror descent. That is, the meta-updates are given as
\begin{IEEEeqnarray}{c}\label{eq:basic_meta_update_u}
	\bm{u}\gets \bm{u}+\bar{\eta}\nabla_{\bm{u}}\phi(\bm{u},\pmb{\lambda})
\end{IEEEeqnarray}
and
\begin{IEEEeqnarray}{c}\label{eq:basic_meta_update_lambda}
	\lambda_m\gets \frac{\lambda_m\exp\rb{\gamma[\nabla_{\pmb{\lambda}} \phi(\bm{u},\pmb{\lambda})]_m}}{\sum_{m'=1}^{M}\lambda_{m'}\exp\rb{\gamma[\nabla_{\pmb{\lambda}} \phi(\bm{u},\pmb{\lambda})]_{m'}}},~\forall m\in[M].
\end{IEEEeqnarray}
In practice, to evaluate the gradients $\nabla_{\bm{u}}\phi(\bm{u},\pmb{\lambda})$ and $\nabla_{\pmb{\lambda}} \phi(\bm{u},\pmb{\lambda})$ of the objective $\phi(\bm{u},\pmb{\lambda})$ in \eqref{eq:opt_meta}, we apply the chain rule of differentiation. Specifically, the meta-update of initialization $\bm{u}$ in \eqref{eq:basic_meta_update_u} can be expressed as
\begin{IEEEeqnarray}{rCl}\label{eq:meta_gd}
	\bm{u}&\gets& \bm{u}+\bar{\eta}\sum_{\tau\in\mathcal T}\mathbf{J}_{\bm{u}_\tau}^\intercal (\bm{u})\nabla_{\bm{u}_\tau}\tilde{R}_\beta^{\mathcal G^\tau}(\exp(\bm{u}_\tau(\bm{u},\pmb{\lambda})),\pmb{\lambda}_\tau(\bm{u},\pmb{\lambda}))\IEEEnonumber\\
	&&+\sum_{\tau\in\mathcal T}\mathbf{J}_{\pmb{\lambda}_\tau}^\intercal (\bm{u})\nabla_{\pmb{\lambda}_\tau}\tilde{R}_\beta^{\mathcal G^\tau}(\exp(\bm{u}_\tau(\bm{u},\pmb{\lambda})),\pmb{\lambda}_\tau(\bm{u},\pmb{\lambda}))\IEEEnonumber\\
	&\triangleq& \text{MGD}_\beta(\bm{u};\mathcal{G}^{\mathcal T},\{\bm{u}_\tau\}_{\tau\in\mathcal T},\{\pmb{\lambda}_\tau\}_{\tau\in\mathcal T}),
\end{IEEEeqnarray}
where $\mathbf{J}_{\bm{f}}^\intercal(\bm{v})$ denotes the transposed Jacobian matrix of vector function $\bm{f}(\bm{v})$, i.e., 
\begin{IEEEeqnarray}{c}
	\mathbf{J}_{\bm{f}}^\intercal(\bm{v})\triangleq\begin{pmatrix}
		\frac{\partial f_1(\bm{v})}{\partial v_1}&\cdots& \frac{\partial f_M(\bm{v})}{\partial v_1}\\
		\vdots&\ddots&\vdots\\
		\frac{\partial f_1(\bm{v})}{\partial v_M}&\cdots& \frac{\partial f_M(\bm{v})}{\partial v_M}
	\end{pmatrix}.
\end{IEEEeqnarray}
Similarly, for $m\in[M]$, the meta-update of initialization $\lambda_m$ in \eqref{eq:basic_meta_update_lambda} can be expressed as
\begin{IEEEeqnarray}{rCl}\label{eq:meta_eg}
	\tilde{\lambda}_m&\gets& \lambda_m\exp\rb{\bar{\gamma}\sqb{\sum_{\tau\in\mathcal T}\mathbf{J}_{\bm{u}_\tau}^\intercal (\pmb{\lambda})\nabla_{\bm{u}_\tau}\tilde{R}_\beta^{\mathcal G^\tau}(\exp(\bm{u}_\tau(\bm{u},\pmb{\lambda})),\pmb{\lambda}_\tau(\bm{u},\pmb{\lambda}))}_m}\IEEEnonumber\\
	&&\cdot \exp\rb{\bar{\gamma}\sqb{\sum_{\tau\in\mathcal T}\mathbf{J}_{\pmb{\lambda}_\tau}^\intercal (\pmb{\lambda})\nabla_{\pmb{\lambda}_\tau}\tilde{R}_\beta^{\mathcal G^\tau}(\exp(\bm{u}_\tau(\bm{u},\pmb{\lambda})),\pmb{\lambda}_\tau(\bm{u},\pmb{\lambda}))}_m}\IEEEnonumber\\
	&\triangleq& \text{MEG}_\beta(\pmb{\lambda};\mathcal{G}^{\mathcal T},\{\bm{u}_\tau\}_{\tau\in\mathcal T},\{\pmb{\lambda}_\tau\}_{\tau\in\mathcal T})
\end{IEEEeqnarray}
with $\lambda_m=\tilde{\lambda}_m/\norms{\tilde{\pmb{\lambda}}}_1$.
In addition, the gradient $\nabla_{\bm{u}_\tau}\tilde{R}_\beta^{\mathcal G^\tau}(\exp(\bm{u}_\tau(\bm{u},\pmb{\lambda})),\pmb{\lambda}_\tau(\bm{u},\pmb{\lambda}))$ in \eqref{eq:meta_gd} and \eqref{eq:meta_eg} can be further expressed as
\begin{IEEEeqnarray}{c}
	\nabla_{\bm{u}_\tau}\tilde{R}_\beta^{\mathcal G^\tau}(\exp(\bm{u}_\tau(\bm{u},\pmb{\lambda})),\pmb{\lambda}_\tau(\bm{u},\pmb{\lambda}))=\diag(\exp(\bm{u}_\tau(\bm{u},\pmb{\lambda})))\left.\nabla_{\bm{s}}\tilde{R}_\beta^{\mathcal G}(\bm{s},\pmb{\lambda}_\tau)\right\rvert_{\bm{s}=\exp(\bm{u}_\tau(\bm{u},\pmb{\lambda}))}.
\end{IEEEeqnarray}
Finally, the \ac{MAML} algorithm is summarized in \cref{alg:maml}.
\begin{algorithm}
	\DontPrintSemicolon
	\caption{MAML}
	\label{alg:maml}
	\SetKwInOut{Input}{Input}
	\Input{Meta-training data $\mathcal{G}^{\mathcal T}$, fraction $\beta\in(0,1]$}
	\KwIn{Randomly initialize $\bm{u}\in\mathbb R^M$ and $\pmb{\lambda}\in\Delta_c^M$}
	\While{not converged}
	{
		\For{each deployment $\tau\in\mathcal{T}$}{
			set $\bm{u}_{\tau}\gets \text{GD}_\beta(\bm{u};\mathcal{G}^\tau,\pmb{\lambda})$  (defined in \eqref{eq:gradient_descent})\;
			set $\pmb{\lambda}_{\tau} \gets \text{EG}_\beta(\pmb{\lambda};\mathcal{G}^\tau,\bm{u})$  (defined in \eqref{eq:exponentiated_gradient}) \;
		}
		
		set $\bm{u}\gets \text{MGD}_\beta(\bm{u};\mathcal{G}^{\mathcal T},\{\bm{u}_\tau\}_{\tau\in\mathcal T},\{\pmb{\lambda}_\tau\}_{\tau\in\mathcal T})$ (defined in \eqref{eq:meta_gd}) \;
		set $\tilde{\lambda}_m\gets \text{MEG}_\beta(\pmb{\lambda};\mathcal{G}^{\mathcal T},\{\bm{u}_\tau\}_{\tau\in\mathcal T},\{\pmb{\lambda}_\tau\}_{\tau\in\mathcal T})$, $\forall m\in[M]$  (defined in \eqref{eq:meta_eg}) \;
		set $\pmb{\lambda}=\tilde{\pmb{\lambda}}/\norms{\tilde{\pmb{\lambda}}}_1$\;
	}
	\Return $\big(\bm{u},\pmb{\lambda}\big)$
\end{algorithm}

\section{Numerical Results}\label{sec:numerical}
In this section, we evaluate the expected rate $\bar{R}\big(\tilde{\bm{s}}_1^{\mathcal G},\tilde{\pmb{\lambda}}_1^{\mathcal G}\big)$ and $\beta$-\ac{CVaR} rate $R_{\beta}\big(\tilde{\bm{s}}_\beta^{\mathcal G},\tilde{\pmb{\lambda}}_\beta^{\mathcal G}\big)$ for parameters $(\tilde{\bm{s}}_1^{\mathcal G},\tilde{\pmb{\lambda}}_1^{\mathcal G})$ and $(\tilde{\bm{s}}_\beta^{\mathcal G},\tilde{\pmb{\lambda}}_\beta^{\mathcal G})$ obtained via Algorithm \ref{alg:empirical_optimization_beta} with learning rates $\eta=\gamma=0.01$ and sigmoid smoothness parameter $c=10$. The expected rate and $\beta$-\ac{CVaR} rate are averaged over 1000 datasets $\mathcal{G}$, which we denote as $\mathbb E_{\mathcal G}\big[\bar{R}\big(\tilde{\bm{s}}_1^{\mathcal G},\tilde{\pmb{\lambda}}_1^{\mathcal G}\big)\big]$, and $\mathbb E_{\mathcal G}\big[R_{\beta}\big(\tilde{\bm{s}}_\beta^{\mathcal G},\tilde{\pmb{\lambda}}_\beta^{\mathcal G}\big)\big]$, respectively. Furthermore, we demonstrate the gain of \ac{MAML} as means to reduce sample complexity by assessing the $\beta$-\ac{CVaR} rate achieved when an initialization $(\bm{s},\pmb{\lambda})$ is optimized based on previous deployments via \cref{alg:maml} with learning rates $\bar{\eta}=\bar{\gamma}=0.01/D$.  
\subsection{Expected Achievable Rate}
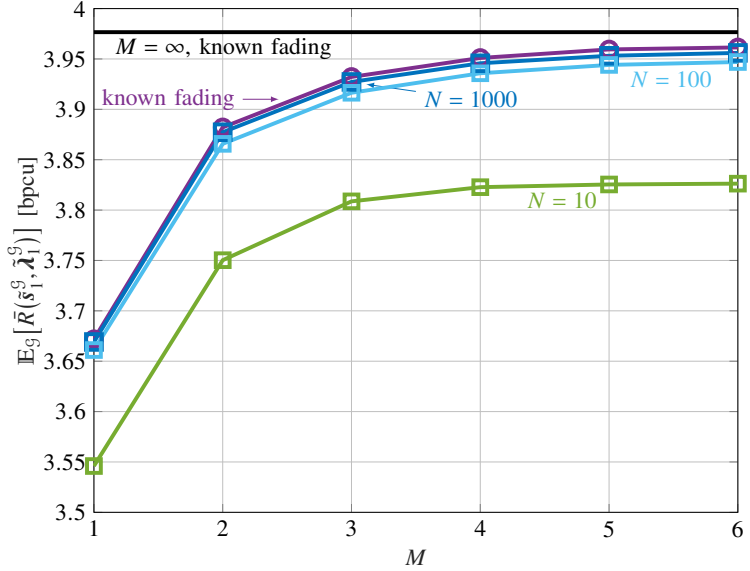
\begin{figure}[!t]
	\centering
	\resizebox {0.7\linewidth} {!} {	
		\definecolor{mycolor1}{rgb}{0.49400,0.18400,0.55600}%
\definecolor{mycolor2}{rgb}{0.00000,0.44700,0.74100}%
\definecolor{mycolor3}{rgb}{0.30100,0.74500,0.93300}%
\definecolor{mycolor4}{rgb}{0.46600,0.67400,0.18800}%
\begin{tikzpicture}

\begin{axis}[%
width=4.521in,
height=3.566in,
at={(0.758in,0.481in)},
scale only axis,
xmin=1,
xmax=6,
xtick={1, 2, 3, 4, 5, 6},
xlabel style={font=\color{white!15!black}},
xlabel={$M$},
ymin=3.5,
ymax=4,
ylabel style={font=\color{white!15!black}},
ylabel={$\mathbb E_{\mathcal G}\big[\bar{R}\big(\tilde{\bm{s}}_1^{\mathcal G},\tilde{\pmb{\lambda}}_1^{\mathcal G}\big)\big]$ [bpcu]},
axis background/.style={fill=white},
xmajorgrids,
ymajorgrids,
legend style={at={(0.97,0.03)}, anchor=south east, legend cell align=left, align=left, draw=white!15!black}
]

\node[anchor=west,color=black] (source) at (axis cs:1.1,3.96){$M=\infty$, known fading};

\node[anchor=west,color=mycolor4] (source2) at (axis cs:4.3,3.81){$N=10$};

\node[anchor=west,color=mycolor1] (source3) at (axis cs:1.0,3.91){known fading};
\node (destination3) at (axis cs:2.5,3.91){};
\draw[-latex,color=mycolor1](source3)--(destination3);

\node[anchor=west,color=mycolor2] (source4) at (axis cs:3.5,3.91){$N=1000$};
\node (destination4) at (axis cs:3.05,3.926){};
\draw[-latex,color=mycolor2](source4)--(destination4);

\node[anchor=west,color=mycolor3] (source5) at (axis cs:5.1,3.93){$N=100$};

\addplot [color=black, line width=2.0pt]
  table[row sep=crcr]{%
1	3.97659973760885\\
2	3.97659973760885\\
3	3.97659973760885\\
4	3.97659973760885\\
5	3.97659973760885\\
6	3.97659973760885\\
};

\addplot [color=mycolor1, line width=2.0pt, mark size=4.0pt, mark=o, mark options={solid, mycolor1}]
  table[row sep=crcr]{%
1	3.67181658464549\\
2	3.88210225678693\\
3	3.93216507080373\\
4	3.95099208877321\\
5	3.95946155695346\\
6	3.96145341793608\\
};

\addplot [color=mycolor2, line width=2.0pt, mark size=4.0pt, mark=square, mark options={solid, mycolor2}]
  table[row sep=crcr]{%
1	3.66908170109552\\
2	3.87728197180082\\
3	3.92718659924209\\
4	3.94571413107387\\
5	3.95316721670574\\
6	3.95600510946281\\
};

\addplot [color=mycolor3, line width=2.0pt, mark size=3.5pt, mark=square, mark options={solid, mycolor3}]
  table[row sep=crcr]{%
1	3.66109036893159\\
2	3.86573467175526\\
3	3.91645006343711\\
4	3.93580427909761\\
5	3.94407467405642\\
6	3.94697163434125\\
};

\addplot [color=mycolor4, line width=2.0pt, mark size=3.5pt, mark=square, mark options={solid, mycolor4}]
  table[row sep=crcr]{%
1	3.5459622217971\\
2	3.75026074261146\\
3	3.8086627285083\\
4	3.82273799119702\\
5	3.82542892478698\\
6	3.82623392895371\\
};

\end{axis}

\begin{axis}[%
width=5.833in,
height=4.375in,
at={(0in,0in)},
scale only axis,
xmin=0,
xmax=1,
ymin=0,
ymax=1,
axis line style={draw=none},
ticks=none,
axis x line*=bottom,
axis y line*=left,
legend style={legend cell align=left, align=left, draw=white!15!black}
]
\end{axis}
\end{tikzpicture}%
	}
	\caption{The expected achievable rate as a function of $M$ with $P=20$dB and $N=10$, $100$, and $1000$.}
	\label{fig:vs_M}
\end{figure}

In \cref{fig:vs_M}, we plot the expected achievable rate as a function of the number of layers $M$ with power $P=20$dB, Rayleigh fading distribution, and dataset of size $N=10$, $100$, and $1000$. 
For this special case, the ideal optimal solution in \cref{cor:shamai} obtained by using infinite layers and assuming that the fading distribution is known is used as an upper bound. Furthermore, we plot for reference the expected rate achieved with finite number of layers when the \ac{BS} knows the fading distribution, which is obtained by replacing the surrogate empirical average rate $\tilde{R}_1^{\mathcal G}(\bm{s},\pmb{\lambda})$ with the expected rate $\bar{R}(\bm{s},\pmb{\lambda})$ in the gradient-based updates  \eqref{eq:gradient_descent}--\eqref{eq:exponentiated_gradient}.
First, confirming the sample complexity analysis in \cref{sec:expected_rate_sample_complexity}, for sufficiently large datasets, the expected rate is close to that achieved when the \ac{BS} knows the fading distribution. Furthermore, using multiple layers provides notable gain over a single layer, even for small datasets. Finally, the expected rate achieved with $M=6$ layers and sufficiently large dataset is seen to be close to the upper bound.

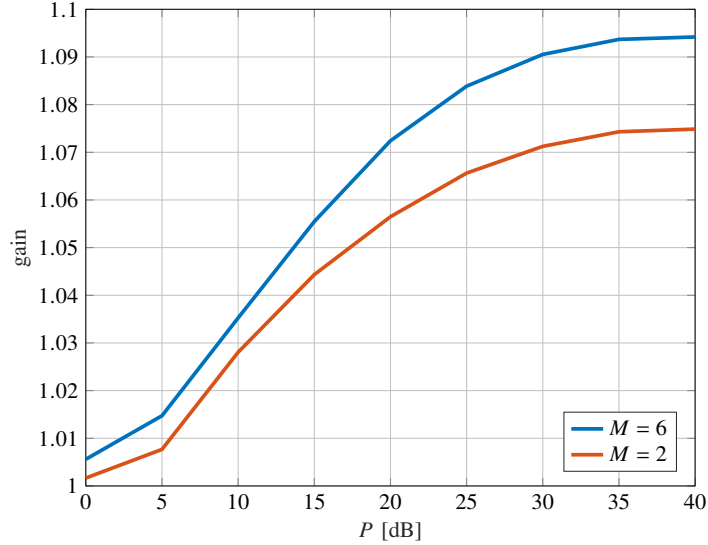
\begin{figure}[!t]
	\centering
	\resizebox {0.6\linewidth} {!} {	
		\definecolor{mycolor1}{rgb}{0.00000,0.44700,0.74100}%
\definecolor{mycolor2}{rgb}{0.85000,0.32500,0.09800}%
\begin{tikzpicture}

\begin{axis}[%
width=4.521in,
height=3.566in,
at={(0.758in,0.481in)},
scale only axis,
xmin=0,
xmax=40,
xlabel style={font=\color{white!15!black}},
xlabel={$P$ [dB]},
ymin=1,
ymax=1.1,
ylabel style={font=\color{white!15!black}},
ylabel={gain},
axis background/.style={fill=white},
xmajorgrids,
ymajorgrids,
legend style={at={(0.97,0.03)}, anchor=south east, legend cell align=left, align=left, draw=white!15!black}
]
\addplot [color=mycolor1, line width=2.0pt]
  table[row sep=crcr]{%
0	1.00559623458653\\
5.00000321342935	1.01472523102312\\
10	1.0352492672367\\
15.0000032134294	1.05547228758428\\
20	1.07241381191062\\
25.0000032134294	1.08386913344571\\
30	1.09053330330517\\
35.0000032134294	1.09368853947913\\
40	1.09419345671631\\
};
\addlegendentry{$M=6$}

\addplot [color=mycolor2, line width=2.0pt]
  table[row sep=crcr]{%
0	1.00165981035277\\
5.00000321342935	1.00766564545378\\
10	1.02805254184254\\
15.0000032134294	1.04433195578668\\
20	1.05648244604552\\
25.0000032134294	1.06563432911598\\
30	1.07124340538494\\
35.0000032134294	1.07430474878696\\
40	1.07486353001281\\
};
\addlegendentry{$M=2$}

\end{axis}
\end{tikzpicture}%
	}
	\caption{The expected achievable rate gain as a function of $\sigma^2$ with $h\sim\mathcal{CN}(0,\sigma^2)$, $P=20$dB, and $N=10^4$.}
	\label{fig:vs_P}
\end{figure}

In \cref{fig:vs_P}, we plot the ratio of the expected rate achieved via \ac{LDM} with $M$ layers to the expected rate achieved with a single layer as a function of the power $P$ with Rayleigh fading distribution and dataset of size $1000$. 
\looseness=-1
It is observed that the gain of \ac{LDM} increases with power $P$. Intuitively, this is because, for sufficiently high power, splitting the last layer, while keeping the same norm $\norms{\bm{s}}_1$, has a negligible impact on the rate $\rho_M(\bm{s},\pmb{\lambda})$ but adds another layer that is much more likely to be decoded (see eqs. \eqref{eq:rate_layer}--\eqref{eq:cvar_expected_rate}). 

\subsection{Conditional Value at Risk}
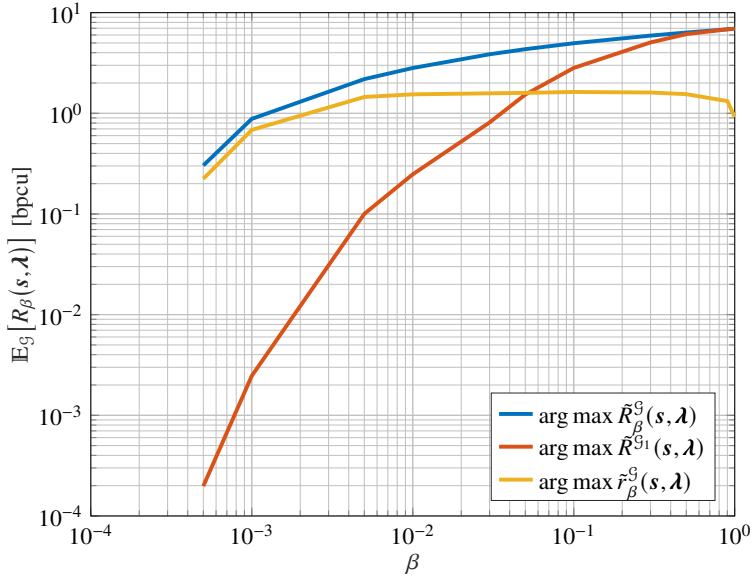
\begin{figure}[!t]
	\centering
	\resizebox {0.7\linewidth} {!} {	
		\definecolor{mycolor1}{rgb}{0.00000,0.44700,0.74100}%
\definecolor{mycolor2}{rgb}{0.85000,0.32500,0.09800}%
\definecolor{mycolor3}{rgb}{0.92900,0.69400,0.12500}%
\begin{tikzpicture}

\begin{axis}[%
width=4.521in,
height=3.563in,
at={(0.758in,0.484in)},
scale only axis,
xmode=log,
xmin=0.0001,
xmax=1,
xminorticks=true,
xlabel style={font=\color{white!15!black}},
xlabel={$\beta$},
ymode=log,
ymin=0.0001,
ymax=10,
yminorticks=true,
ylabel style={font=\color{white!15!black}},
ylabel={$\mathbb E_{\mathcal G}\big[R_{\beta}\big(\bm{s},\pmb{\lambda}\big)\big]$ [bpcu]},
axis background/.style={fill=white},
xmajorgrids,
xminorgrids,
ymajorgrids,
yminorgrids,
legend style={at={(0.97,0.03)}, anchor=south east, legend cell align=left, align=left, draw=white!15!black}
]
\addplot [color=mycolor1, line width=2.0pt]
  table[row sep=crcr]{%
0.0005	0.303445\\
0.001	0.880789\\
0.005	2.18695\\
0.01	2.8195\\
0.03	3.86017\\
0.05	4.33619\\
0.1	4.96576\\
0.3	5.91546\\
0.5	6.34184\\
0.9	6.83209\\
1	6.9215\\
};
\addlegendentry{$\argmax \tilde{R}^\mathcal{G}_\beta(\bm{s},\pmb{\lambda})$}

\addplot [color=mycolor2, line width=2.0pt]
  table[row sep=crcr]{%
0.0005	0.000198048\\
0.001	0.00246536\\
0.005	0.100203\\
0.01	0.247691\\
0.03	0.811701\\
0.05	1.53443\\
0.1	2.82246\\
0.3	5.06373\\
0.5	6.11682\\
0.9	6.83157\\
1	6.9215\\
};
\addlegendentry{$\argmax \tilde{R}^{\mathcal{G}_1}(\bm{s},\pmb{\lambda})$}

\addplot [color=mycolor3, line width=2.0pt]
  table[row sep=crcr]{%
0.0005	0.223565\\
0.001	0.684231\\
0.005	1.45302\\
0.01	1.54354\\
0.03	1.57956\\
0.05	1.59615\\
0.1	1.62947\\
0.3	1.6095\\
0.5	1.55022\\
0.9	1.32281\\
1	0.9164\\
};
\addlegendentry{$\argmax \tilde{r}^\mathcal{G}_\beta(\bm{s},\pmb{\lambda})$}

\end{axis}

\begin{axis}[%
width=5.833in,
height=4.375in,
at={(0in,0in)},
scale only axis,
xmin=0,
xmax=1,
ymin=0,
ymax=1,
axis line style={draw=none},
ticks=none,
axis x line*=bottom,
axis y line*=left,
legend style={legend cell align=left, align=left, draw=white!15!black}
]
\end{axis}
\end{tikzpicture}%
	}
	\caption{The expected $\beta$-\ac{CVaR} rate as a function of $\beta$ with $P=20$dB, $M=6$, and $N=10^4$.}
	\label{fig:vs_beta}
\end{figure}

To demonstrate the importance of optimizing the $\beta$-\ac{CVaR} rate for applications that focus on the performance of a small fraction $\beta$ of the clients with the worst channel gains, in \cref{fig:vs_beta}, we plot the expected $\beta$-\ac{CVaR} rate as a function of the fraction $\beta$ with power $P=20$dB, Rician fading distribution $h\sim\mathcal{CN}(2,1)$, $M=6$ layers, and datasets of size $N=10^4$, and for rate- and power-allocation vectors $\bm{s}$ and $\pmb{\lambda}$ optimized based on different metrics: \emph{(i)} the surrogate empirical $\beta$-\ac{CVaR} rate $\tilde{R}_\beta^{\mathcal{G}}(\bm{s},\pmb{\lambda})$ defined in \eqref{eq:explicit_surrogate_empirical_cvar}, \emph{(ii)} the surrogate average rate $\tilde{R}^{\mathcal G_1}(\bm{s},\pmb{\lambda})$ defined in \eqref{eq:surrogate_obj_1}, and \emph{(iii)} the surrogate empirical outage rate $\tilde{r}^{\mathcal{G}}_\beta(\bm{s},\pmb{\lambda})=R_\sigma(\bm{s},\pmb{\lambda},g_{[\nint{N\beta}]})$.
It is observed that, for most values of $\beta\in(0,1]$, the average achievable rate for the $\beta$-fraction of  clients with the worst channel gains increases significantly if the rate- and power-allocation vectors are optimized to maximize the surrogate empirical $\beta$-\ac{CVaR} rate. In contrast, for sufficiently high fraction $\beta$, the surrogate average rate can also be used as the optimization objective. This is because the surrogate average rate $\tilde{R}^{\mathcal G_1}(\bm{s},\pmb{\lambda})$ is a special case of the surrogate empirical $\beta$-\ac{CVaR} rate $\tilde{R}_\beta^{\mathcal{G}}(\bm{s},\pmb{\lambda})$ for $\beta=1$. Similarly, for sufficiently low fraction $\beta$, the surrogate empirical outage rate can used as the optimization objective. This is due to the limit $\lim_{\beta\to0}\tilde{R}_\beta^{\mathcal{G}}(\bm{s},\pmb{\lambda})=\lim_{\beta\to0}\tilde{r}^{\mathcal{G}}_\beta(\bm{s},\pmb{\lambda})=R_\sigma(\bm{s},\pmb{\lambda},g_{[1]})$ (see eq. \eqref{eq:explicit_surrogate_empirical_cvar}).

\begin{figure}[!t]
	\centering
	\resizebox {0.6\linewidth} {!} {	
		\definecolor{mycolor1}{rgb}{0.85000,0.32500,0.09800}%
\definecolor{mycolor2}{rgb}{0.46600,0.67400,0.18800}%
\definecolor{mycolor3}{rgb}{0.00000,0.44700,0.74100}%
\begin{tikzpicture}

\begin{axis}[%
width=4.521in,
height=3.563in,
at={(0.758in,0.484in)},
scale only axis,
xmode=log,
xmin=1,
xmax=1000,
xminorticks=true,
xlabel style={font=\color{white!15!black}},
xlabel={$N$},
ymin=0,
ymax=9,
ylabel style={font=\color{white!15!black}},
ylabel={$\mathbb E_{\mathcal G}\big[R_{\beta}\big(\tilde{\bm{s}}_\beta^{\mathcal G},\tilde{\pmb{\lambda}}_\beta^{\mathcal G}\big)\big]$ [bpcu]},
axis background/.style={fill=white},
xmajorgrids,
xminorgrids,
ymajorgrids,
legend style={at={(0.03,0.5)}, anchor=west, legend cell align=left, align=left, draw=white!15!black}
]

\node (A) at (axis cs:15,8.25){};
\draw[thick,draw=black] (A) ellipse (5pt and 20pt);
\node at (axis cs:15,7.3){$\beta=1$};
\node (B) at (axis cs:15,3.4){};
\draw[thick,draw=black] (B) ellipse (5pt and 15pt);
\node at (axis cs:15,2.4){$\beta=0.1$};
\node (C) at (axis cs:150,2.3){};
\draw[thick,draw=black] (C) ellipse (5pt and 15pt);
\node at (axis cs:150,1.4){$\beta=0.01$};

\addplot [color=mycolor1, dashed, line width=2.0pt]
  table[row sep=crcr]{%
1	0.315763511168175\\
2	0.695814063395972\\
3	1.04066039419376\\
4	1.37564200343004\\
5	1.62919198119221\\
6	1.9281884919902\\
7	2.19241542844661\\
8	2.40097448058389\\
10	2.72905130100798\\
11	2.89754892376197\\
13	3.1702888014294\\
15	3.29248428571545\\
17	3.4822401026924\\
19	3.67902877956807\\
22	3.88764588172695\\
26	4.04570245677357\\
29	4.18137968833907\\
34	4.42273886852643\\
39	4.49348638735187\\
45	4.61249465545552\\
52	4.74368795733596\\
60	4.82462222620113\\
69	4.88992674334046\\
79	4.9750744914631\\
91	5.05306980672869\\
105	5.11996387521466\\
121	5.14544661822637\\
139	5.19888883649877\\
160	5.22426026763018\\
184	5.27070610798304\\
212	5.29891325680194\\
244	5.31916639421316\\
281	5.32817729364224\\
324	5.34387861379906\\
373	5.36582157615752\\
429	5.37859707702295\\
494	5.38633062893705\\
569	5.39546239416812\\
655	5.40253138676016\\
754	5.40805916140765\\
869	5.41513668823846\\
1000	5.42027184984804\\
};
\addlegendentry{$M=1$}

\addplot [color=mycolor2, dotted, line width=2.0pt]
  table[row sep=crcr]{%
1	0.349369815945504\\
2	0.779942364128944\\
3	1.15523281165168\\
4	1.51044605968696\\
5	1.78373775384189\\
6	2.10227746970256\\
7	2.3787937562202\\
8	2.59833760323444\\
10	2.94256759593962\\
11	3.11095539194349\\
13	3.38508371450105\\
15	3.6113664205436\\
17	3.79422481915863\\
19	4.00109037558851\\
22	4.1748455306861\\
26	4.3388693797282\\
29	4.47886481239745\\
34	4.72157675735899\\
39	4.79334214455508\\
45	4.8699066085472\\
52	5.01950783634224\\
60	5.10207397120554\\
69	5.18749230464771\\
79	5.26963645906487\\
91	5.36054161660803\\
105	5.39844749164179\\
121	5.45942475151287\\
139	5.49557006377814\\
160	5.52826097078958\\
184	5.56607648457048\\
212	5.58013010746543\\
244	5.60558146089518\\
281	5.61766247789176\\
324	5.63321847307557\\
373	5.65092767105035\\
429	5.65942712200391\\
494	5.67591763970272\\
569	5.68065221170885\\
655	5.68804373363863\\
754	5.6981632837732\\
869	5.70476154051218\\
1000	5.71118643179459\\
};
\addlegendentry{$M=2$}

\addplot [color=mycolor3, line width=2.0pt]
  table[row sep=crcr]{%
1	0.334611135494425\\
2	0.740392564294976\\
3	1.09903616933657\\
4	1.44311426712674\\
5	1.70904784038524\\
6	2.02722097214373\\
7	2.30140155779536\\
8	2.52118651792297\\
10	2.86540774641799\\
11	3.03693732689424\\
13	3.31732978825695\\
15	3.57633139691932\\
17	3.78028904497517\\
19	3.98018962678435\\
22	4.19077148788058\\
26	4.37467559282026\\
29	4.54716193370042\\
34	4.78546164159558\\
39	4.86972281417364\\
45	4.97255813944943\\
52	5.13386591881808\\
60	5.21197813489234\\
69	5.28448918609384\\
79	5.36485185381167\\
91	5.44957479160158\\
105	5.49470492159286\\
121	5.56184531463769\\
139	5.60394374108072\\
160	5.6391899157728\\
184	5.67920958898288\\
212	5.70159918514711\\
244	5.72591734075054\\
281	5.73966607206743\\
324	5.75771502954834\\
373	5.77724256377053\\
429	5.78937590197635\\
494	5.80202923350995\\
569	5.81113955108292\\
655	5.81974462744278\\
754	5.82952661448683\\
869	5.83379973579125\\
1000	5.8397062225919\\
};
\addlegendentry{$M=6$}

\addplot [color=mycolor1, dashed, line width=2.0pt, forget plot]
  table[row sep=crcr]{%
1	7.4159950365873\\
2	7.60681361621743\\
3	7.71998675950312\\
4	7.78766774405408\\
5	7.83203400452475\\
6	7.86587017864315\\
7	7.8951184811685\\
8	7.91855996641225\\
10	7.97726404625595\\
11	7.99668541188272\\
13	8.03661147425297\\
15	8.05818854717878\\
17	8.07987525209492\\
19	8.09702835605505\\
22	8.11659484515169\\
26	8.13432018132696\\
29	8.1381694439706\\
34	8.1521189329941\\
39	8.15891733502004\\
45	8.17057777130402\\
52	8.17464864920349\\
60	8.18190992255208\\
69	8.18768977889683\\
79	8.19103978589671\\
91	8.19325543214875\\
105	8.19765173076124\\
121	8.19887070812801\\
139	8.20108309162965\\
160	8.20033912703827\\
184	8.20367904330928\\
212	8.20583693604431\\
244	8.20701192001856\\
281	8.20852697554859\\
324	8.21054729494381\\
373	8.21261589096609\\
429	8.21303768789325\\
494	8.21446743013909\\
569	8.21525933935021\\
655	8.21671536105776\\
754	8.21736421790203\\
869	8.21816004573229\\
1000	8.2187843513755\\
};
\addplot [color=mycolor2, dotted, line width=2.0pt, forget plot]
  table[row sep=crcr]{%
1	6.95664137295266\\
2	7.43142207522109\\
3	7.75714482371161\\
4	7.96024273368552\\
5	8.08290667448508\\
6	8.18705326202633\\
7	8.23867530321712\\
8	8.3033448920299\\
10	8.39115256939342\\
11	8.40938141401796\\
13	8.44846356325064\\
15	8.4723516696731\\
17	8.48331786764997\\
19	8.49173582770534\\
22	8.50441804662522\\
26	8.50858896929923\\
29	8.50534097624737\\
34	8.50131548516467\\
39	8.5034927924973\\
45	8.51253371800673\\
52	8.52525056916237\\
60	8.53896150154417\\
69	8.55271940318042\\
79	8.56263072663685\\
91	8.56933931675331\\
105	8.58023643200566\\
121	8.58835143607761\\
139	8.599854224849\\
160	8.6095541037587\\
184	8.61717470407731\\
212	8.63261083378139\\
244	8.64098630744823\\
281	8.65094434365179\\
324	8.66213875551733\\
373	8.66951475938884\\
429	8.67642194113609\\
494	8.68296056603435\\
569	8.69158320877139\\
655	8.69483500825634\\
754	8.70026632466878\\
869	8.70432372640403\\
1000	8.70914492693313\\
};
\addplot [color=mycolor3, line width=2.0pt, forget plot]
  table[row sep=crcr]{%
1	5.9248213042048\\
2	7.06775433014458\\
3	7.66183535382796\\
4	7.96612397104425\\
5	8.14674911498295\\
6	8.28435008247376\\
7	8.36471833803667\\
8	8.43481430612689\\
10	8.53357898626162\\
11	8.56111872491854\\
13	8.60946823632722\\
15	8.64977913016857\\
17	8.67136043744262\\
19	8.6837277323773\\
22	8.69853523243451\\
26	8.71524308431802\\
29	8.72353747055091\\
34	8.73234573115368\\
39	8.73751285078781\\
45	8.74470982180963\\
52	8.75229761450496\\
60	8.75782675432422\\
69	8.76632070051233\\
79	8.77387869304238\\
91	8.78148759742941\\
105	8.78788903294601\\
121	8.79339032902583\\
139	8.79876287377166\\
160	8.80340705494562\\
184	8.80932991235505\\
212	8.81361112495698\\
244	8.81826468033458\\
281	8.82523667726602\\
324	8.82773711976229\\
373	8.83097825163307\\
429	8.83438263896418\\
494	8.83744723725399\\
569	8.84015600521538\\
655	8.84279786522127\\
754	8.84468066608384\\
869	8.84658711331359\\
1000	8.84838757190985\\
};
\addplot [color=mycolor1, dashed, line width=2.0pt, forget plot]
  table[row sep=crcr]{%
1	0.00115551039268574\\
2	0.0218831688129856\\
3	0.0463838099365735\\
4	0.102924944023115\\
5	0.134434432833258\\
7	0.206578040467577\\
9	0.270383368851765\\
11	0.299461361409679\\
14	0.359856238711806\\
17	0.428515829736259\\
22	0.616389567157554\\
28	0.789742603577845\\
36	0.975004711922739\\
45	1.16908726218067\\
57	1.34013619770494\\
73	1.57282300993377\\
92	1.84318877581181\\
117	2.08431545435459\\
149	2.32354022663195\\
189	2.38576548315295\\
240	2.52308866558785\\
304	2.55595542067596\\
386	2.64529800668758\\
489	2.71744960281094\\
621	2.78121222728553\\
788	2.80838259840906\\
1000	2.84026925738137\\
};
\addplot [color=mycolor2, dotted, line width=2.0pt, forget plot]
  table[row sep=crcr]{%
1	0.00832116584909727\\
2	0.0470650184652383\\
3	0.0613544694071737\\
4	0.105685733130485\\
5	0.153262927569321\\
7	0.211567759145813\\
9	0.327740994428959\\
11	0.370168384417628\\
14	0.42629050968072\\
17	0.494457112973673\\
22	0.662707738045373\\
28	0.845272463691651\\
36	1.04480578444552\\
45	1.25945161146377\\
57	1.44275250920457\\
73	1.68490960428615\\
92	1.96145512882705\\
117	2.2069481449489\\
149	2.44527251528334\\
189	2.47398198471887\\
240	2.63339499809412\\
304	2.68201391653891\\
386	2.76594087542657\\
489	2.83656087604854\\
621	2.89798492475509\\
788	2.91913882190485\\
1000	2.951427825552\\
};
\addplot [color=mycolor3, line width=2.0pt, forget plot]
  table[row sep=crcr]{%
1	0.0208381255929029\\
2	0.0657956623327752\\
3	0.0868146735626865\\
4	0.130917833238366\\
5	0.15429568407997\\
7	0.223219418420164\\
9	0.302836203665495\\
11	0.364859388756477\\
14	0.428055362229984\\
17	0.516619574407473\\
22	0.645614278808847\\
28	0.829528618339931\\
36	1.06445674896483\\
45	1.27673491174365\\
57	1.4717790849794\\
73	1.72138246258776\\
92	2.00309463121027\\
117	2.2487997981525\\
149	2.48437218445304\\
189	2.51932386486713\\
240	2.67622364322183\\
304	2.72330488967325\\
386	2.80704067001478\\
489	2.873103309392\\
621	2.93323249875865\\
788	2.95239945201652\\
1000	2.98456116753404\\
};
\end{axis}
\end{tikzpicture}%
	}
	\caption{The expected $\beta$-\ac{CVaR} rate as a function of $N$ with $P=20$dB, $M=1$, $2$, and $6$, and $\beta=1$, $0.1$, and $0.01$.}
	\label{fig:cvar_vs_N}
\end{figure}
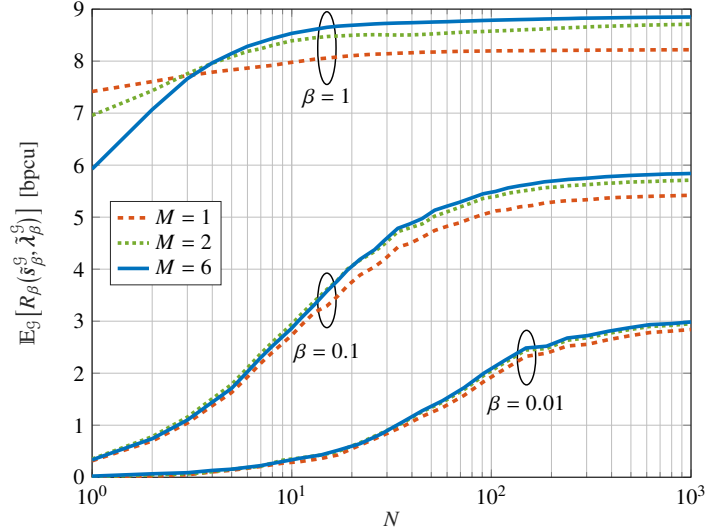

In \cref{fig:cvar_vs_N}, we plot the expected $\beta$-\ac{CVaR} rate, $\mathbb E_{\mathcal G}\big[R_{\beta}\big(\tilde{\bm{s}}_\beta^{\mathcal G},\tilde{\pmb{\lambda}}_\beta^{\mathcal G}\big)\big]$, as a function of the dataset size $N$ with power $P=20$dB, Rician fading distribution $h\sim\mathcal{CN}(\sqrt{20},16)$, $M=1$, $2$, and $6$ layers, and for $\beta=1$, $0.1$, and $0.01$.
It is observed that, as $\beta$ decreases, larger dataset is required to obtain an accurate estimate of the achievable $\beta$-\ac{CVaR} rate. This is because, as discussed in \cref{sec:CVaR_sample_complexity}, the number of samples required to maintain a desired optimality gap increases as $\beta\rightarrow 0$. In addition, using multiple layers is shown to be advantageous even for very small datasets ($N\geq 4$).

\subsection{Meta-Learning}
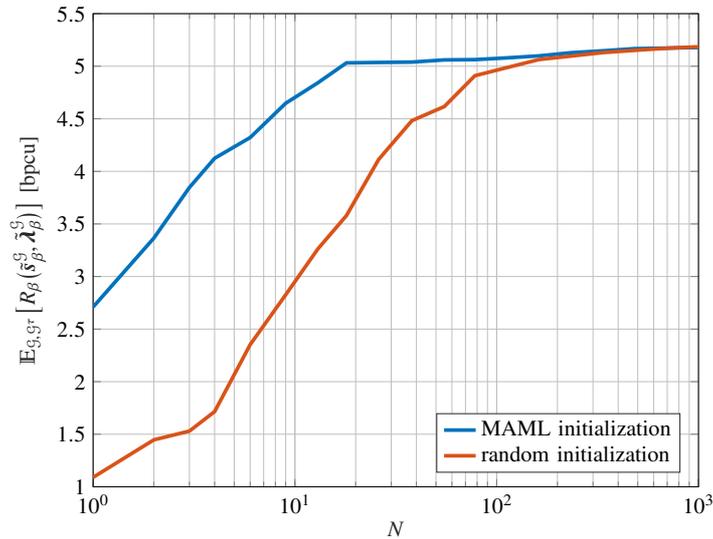
\begin{figure}[!t]
	\centering
	\resizebox {0.6\linewidth} {!} {	
		\definecolor{mycolor1}{rgb}{0.00000,0.44700,0.74100}%
\definecolor{mycolor2}{rgb}{0.85000,0.32500,0.09800}%
\begin{tikzpicture}

\begin{axis}[%
width=4.521in,
height=3.563in,
at={(0.758in,0.484in)},
scale only axis,
xmode=log,
xmin=1,
xmax=1000,
xminorticks=true,
xlabel style={font=\color{white!15!black}},
xlabel={$N$},
ymin=1,
ymax=5.5,
ylabel style={font=\color{white!15!black}},
ylabel={$\mathbb E_{\mathcal G,\mathcal G^\tau}\big[R_{\beta}\big(\tilde{\bm{s}}_\beta^{\mathcal G},\tilde{\pmb{\lambda}}_\beta^{\mathcal G}\big)\big]$ [bpcu]},
axis background/.style={fill=white},
xmajorgrids,
xminorgrids,
ymajorgrids,
legend style={at={(0.97,0.03)}, anchor=south east, legend cell align=left, align=left, draw=white!15!black}
]
\addplot [color=mycolor1, line width=2.0pt]
  table[row sep=crcr]{%
1	2.71038\\
2	3.36547\\
3	3.84656\\
4	4.1244\\
6	4.32139\\
9	4.64751\\
13	4.84252\\
18	5.03169\\
26	5.0358\\
38	5.0392\\
55	5.06\\
78	5.06269\\
113	5.0796\\
162	5.09844\\
234	5.12851\\
336	5.14668\\
483	5.16734\\
695	5.17238\\
1000	5.17793\\
};
\addlegendentry{MAML initialization}

\addplot [color=mycolor2, line width=2.0pt]
  table[row sep=crcr]{%
1	1.09024\\
2	1.44652\\
3	1.5298\\
4	1.7154\\
6	2.35222\\
9	2.82618\\
13	3.26274\\
18	3.57912\\
26	4.11339\\
38	4.48255\\
55	4.61631\\
78	4.90975\\
113	4.98814\\
162	5.06334\\
234	5.09667\\
336	5.12979\\
483	5.15092\\
695	5.17103\\
1000	5.18504\\
};
\addlegendentry{random initialization}

\end{axis}
\end{tikzpicture}%
}
	\caption{The expected $\beta$-\ac{CVaR} rate as a function of $N$ with $P=20$dB, $M=6$, $\beta=0.1$, and $D=10$.}
	\label{fig:maml_vs_N}
\end{figure}

To evaluate the gain of meta-learning in reducing the sample complexity, we compare the expected $\beta$-\ac{CVaR} rate achieved with random initialization to that achieved by optimizing an initialization $(\bm{s},\pmb{\lambda})$ based on the data from previous deployments via \cref{alg:maml}. To this end, in \cref{fig:maml_vs_N}, we plot the expected $\beta$-\ac{CVaR} rate as a function of the dataset size $N$ with power $P=20$dB, $M=6$ layers, $\beta=0.1$, and $D=10$ previous deployments. The channel for the new deployment is taken to be $h\sim\mathcal{CN}(\sqrt{10},5)$, whereas the channel for each previous deployment is given as $h_\tau\sim\mathcal{CN}(\sqrt{10}+\mu_\tau,5)$ with random deviation $\mu_\tau\sim\mathcal{CN}(0,2)$.
For small datasets, it is observed that \ac{MAML} provides significant performance gains over the $\beta$-\ac{CVaR} rate achieved with random initialization, i.e., when the power and rate allocation vectors are optimized without taking into account data from previous deployments. In contrast, for sufficiently large datasets, local optimization for each new deployment is sufficient for achieving a desired optimality gap.  

\begin{figure}[!t]
	\centering
	\resizebox {0.6\linewidth} {!} {	
		\definecolor{mycolor1}{rgb}{0.00000,0.44700,0.74100}%
\definecolor{mycolor2}{rgb}{0.85000,0.32500,0.09800}%
\begin{tikzpicture}

\begin{axis}[%
width=4.521in,
height=3.566in,
at={(0.758in,0.481in)},
scale only axis,
xmin=2,
xmax=26,
xtick={ 2,  4,  6,  8, 10, 12, 14, 16, 18, 20, 22, 24, 26},
xlabel style={font=\color{white!15!black}},
xlabel={$D$},
ymin=2.5,
ymax=5,
ylabel style={font=\color{white!15!black}},
ylabel={$\mathbb E_{\mathcal G,\mathcal G^\tau}\big[R_{\beta}\big(\tilde{\bm{s}}_\beta^{\mathcal G},\tilde{\pmb{\lambda}}_\beta^{\mathcal G}\big)\big]$ [bpcu]},
axis background/.style={fill=white},
xmajorgrids,
ymajorgrids,
legend style={at={(0.97,0.03)}, anchor=south east, legend cell align=left, align=left, draw=white!15!black}
]
\addplot [color=mycolor1, line width=2.0pt]
  table[row sep=crcr]{%
2	3.52966333333333\\
4	4.39264\\
6	4.64700333333333\\
8	4.70099333333333\\
10	4.72210333333333\\
12	4.80202\\
14	4.85441666666667\\
16	4.93834\\
18	4.95356\\
20	4.95963\\
22	4.98846666666667\\
24	4.99519666666667\\
26	4.99468333333333\\
};
\addlegendentry{MAML initialization}

\addplot [color=mycolor2, line width=2.0pt]
  table[row sep=crcr]{%
2	2.9\\
4	2.9\\
6	2.9\\
8	2.9\\
10	2.9\\
12	2.9\\
14	2.9\\
16	2.9\\
18	2.9\\
20	2.9\\
22	2.9\\
24	2.9\\
26	2.9\\
};
\addlegendentry{random initialization}

\end{axis}

\begin{axis}[%
width=5.833in,
height=4.375in,
at={(0in,0in)},
scale only axis,
xmin=0,
xmax=1,
ymin=0,
ymax=1,
axis line style={draw=none},
ticks=none,
axis x line*=bottom,
axis y line*=left,
legend style={legend cell align=left, align=left, draw=white!15!black}
]
\end{axis}
\end{tikzpicture}%
}
	\caption{The expected $\beta$-\ac{CVaR} rate as a function of $D$ with $P=20$dB, $M=6$, $\beta=0.1$, and $N=10$.}
	\label{fig:maml_vs_D}
\end{figure}
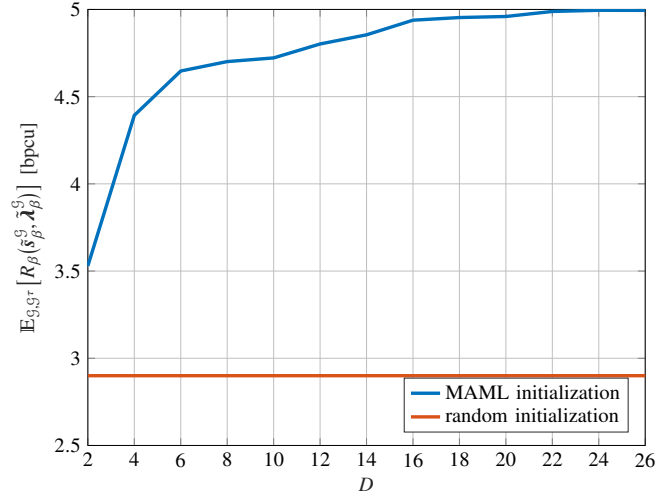

In \cref{fig:maml_vs_D}, we plot the expected $\beta$-\ac{CVaR} rate as a function of the number of previous deployments $D$ with power $P=20$dB, $M=6$ layers, $\beta=0.1$, datasets of size $N=10$, and channel coefficients $\mathrm{h}$ and $\{\mathrm{h}_\tau\}_\tau\in\mathcal{T}$ distributed as in the previous numerical experiment.
While the expected $\beta$-\ac{CVaR} rate is monotonically increasing with the number of previous deployments, \cref{fig:maml_vs_D} demonstrates that the sample complexity can be reduced substantially even if \ac{MAML} is applied with only a few deployments.

\section{Conclusion}\label{sec:conclusion}
In this work, we have studied \ac{LDM} as an enabler of differential \ac{QoS} for ultra-reliable broadcast/multicast communication systems. To this end, we have introduced the $\beta$-\ac{CVaR} rate as the average rate of the $\beta$-fraction of clients with the worst instantaneous channels. We have focused on a practical model in which the fading distribution is unknown, and the transmitter optimizes rate and power allocation for each layer based on a dataset sampled during deployment. The optimality gap caused by the availability of limited data was bounded via a generalization analysis, and the sample complexity was shown to increase as the fraction $\beta$ decreases. To optimize the rate and power allocation parameters, a mirror-descent based scheme was introduced, which, for sufficiently large dataset, was demonstrated via numerical experiments to achieve a $\beta$-\ac{CVaR} rate close to that achieved when the \ac{BS} known the fading distribution. Furthermore, meta-learning was shown to be instrumental in decreasing the sample complexity for ultra-reliable communication with low outage probabilities. Among related problems left open by this study, we mention the extension to multiple transmit antennas \cite{shi2018learning}, to channels with multiple uncoordinated transmitters \cite{zohdy2019broadcast,zohdy2021distributed}, and the analysis of \ac{LDM} with an infinite number of layers \cite{shamai2003broadcast}.  

\appendix
\subsection{Proof of Proposition \ref{prop:expected_rate_convergence}}\label{app:proof_prop_expected_rate_convergence}
The optimality gap $e^{\mathcal G}$ \eqref{eq:gen_error} can be upper bounded as
\begin{IEEEeqnarray}{rCl}\label{eq:basic_gen_ineq}
	e^{\mathcal G}&=&\bar{R}(\bm{s}^*,\pmb{\lambda}^*)-\bar{R}^{\mathcal G}(\bm{s}^*,\pmb{\lambda}^*)+\bar{R}^{\mathcal G}(\bm{s}^*,\pmb{\lambda}^*)-\bar{R}^{\mathcal G}(\bm{s}^{\mathcal G},\pmb{\lambda}^{\mathcal G})+\bar{R}^{\mathcal G}(\bm{s}^{\mathcal G},\pmb{\lambda}^{\mathcal G})-\bar{R}(\bm{s}^{\mathcal G},\pmb{\lambda}^{\mathcal G})\IEEEnonumber\\
	&\leq&\rb{\bar{R}(\bm{s}^*,\pmb{\lambda}^*)-\bar{R}^{\mathcal G}(\bm{s}^*,\pmb{\lambda}^*)}+\rb{\bar{R}^{\mathcal G}(\bm{s}^{\mathcal G},\pmb{\lambda}^{\mathcal G})-\bar{R}(\bm{s}^{\mathcal G},\pmb{\lambda}^{\mathcal G})},
\end{IEEEeqnarray}
where the inequality holds since $(\bm{s}^{\mathcal G},\pmb{\lambda}^{\mathcal G})$ maximize the average rate $\bar{R}^{\mathcal G}(\bm{s},\pmb{\lambda})$. Next, to further bound the optimality gap, we bound, uniformly, the difference $|\bar{R}(\bm{s},\pmb{\lambda})-\bar{R}^{\mathcal G}(\bm{s},\pmb{\lambda})|$ for all $\pmb{\lambda}\in\Delta_c^M$ and $\bm{s}\in\mathbb R_+^M$ with $\norms{\bm{s}}_1\leq S$.
Note that the expected achievable rate \eqref{eq:cvar_expected_rate} can be expressed as
\begin{IEEEeqnarray}{rCl}\label{excpected_rate}
	\bar{R}(\bm{s},\pmb{\lambda})&=&\mathbb E_{\mathrm g}\sqb{R(\bm{s},\pmb{\lambda},\mathrm{g})}
	=\sum_{m=1}^{M}\rho_m(\bm{s}^m,\pmb{\lambda})\bar{F}_{\mathrm{g}}(\norms{\bm{s}^m}_1),
\end{IEEEeqnarray}
where $\bar{F}_{\mathrm{g}}(\norms{\bm{s}^m}_1)$ denotes the \ac{CCDF}
\begin{IEEEeqnarray}{c}
	\bar{F}_{\mathrm{g}}(\norms{\bm{s}^m}_1)\triangleq \Pr\sqb{\mathrm{g}\geq \norms{\bm{s}^m}_1}.
\end{IEEEeqnarray}
Similarly, the average rate \eqref{eq:empir_exp_rate} can be expressed as
\begin{IEEEeqnarray}{rCl}\label{empirical_excpected_rate}
	\bar{R}^{\mathcal G}(\bm{s},\pmb{\lambda})&=&\frac{1}{N}\sum_{i=1}^{N}R(\bm{s},\pmb{\lambda},g_i)=\frac{1}{N}\sum_{i=1}^{N}\sum_{m=1}^{M}\rho_m(\bm{s}^m,\pmb{\lambda})\bm{1}_{g_i\geq \norms{\bm{s}^m}_1}\IEEEnonumber\\
	&=&\frac{1}{N}\sum_{m=1}^{M}\rho_m(\bm{s}^m,\pmb{\lambda})\sum_{i=1}^{N}\bm{1}_{g_i\geq \norms{\bm{s}^m}_1}\IEEEnonumber\\
	&=&\sum_{m=1}^{M}\rho_m(\bm{s}^m,\pmb{\lambda})\bar{F}^{\mathcal G}_{\mathrm{g}}(\norms{\bm{s}^m}_1),
\end{IEEEeqnarray}
where $\bar{F}^{\mathcal G}_{\mathrm{g}}(\norms{\bm{s}^m}_1)$ denotes the empirical \ac{CCDF} 
\begin{IEEEeqnarray}{c}\label{eq:cdf_approx}
	\bar{F}^{\mathcal G}_{\mathrm{g}}(\norms{\bm{s}^m}_1)\triangleq \frac{1}{N}\sum_{i=1}^{N}\bm{1}_{g_i\geq \norms{\bm{s}^m}_1}.
\end{IEEEeqnarray}
Therefore, to uniformly bound the difference $|\bar{R}(\bm{s},\pmb{\lambda})-\bar{R}^{\mathcal G}(\bm{s},\pmb{\lambda})|$, we first uniformly bound $|\bar{F}_{\mathrm{g}}(s)-\bar{F}^{\mathcal G}_{\mathrm{g}}(s)|$ using the following proposition.
\begin{proposition}\label{prop:CCDF_convergence}
	Let $\mathcal G=\{g_1,\ldots,g_N\}$ be a dataset of $N$ fading realizations drawn independently from the fading distribution $p_{\mathrm{g}}(g)$, and let $\delta\in(0,1]$. With probability at least $1-\delta$, uniformly over all $s\in\mathbb R_+$, we have
	\begin{IEEEeqnarray}{c}\label{eq:prop_convergence}
		\abs{\bar{F}_{\mathrm{g}}(s)-\bar{F}^{\mathcal G}_{\mathrm{g}}(s)}\leq 4\sqrt{\frac{(2N+1)\ln(N+1)}{3N(N+1)}}+\sqrt{\frac{2\ln(2/\delta)}{N}}.
	\end{IEEEeqnarray}
\end{proposition}
\begin{IEEEproof}
	See Appendix \ref{app:proof_ccdf_conv}.
\end{IEEEproof}

\cref{prop:CCDF_convergence} implies that, with probability at least $1-\delta$, we can bound the difference  $|\bar{R}(\bm{s},\pmb{\lambda})-\bar{R}^{\mathcal G}(\bm{s},\pmb{\lambda})|$, uniformly over all $\pmb{\lambda}\in\Delta_c^M$ and $\bm{s}\in\mathbb R_+^M$ with $\norms{\bm{s}}_1\leq S$, as
\begin{IEEEeqnarray}{rCl}\label{eq:rate_conv}
	\abs{\bar{R}(\bm{s},\pmb{\lambda})-\bar{R}^{\mathcal G}(\bm{s},\pmb{\lambda})}
	&\overset{\text{(a)}}{=}&\abs{\sum_{m=1}^{M}\rho_m(\bm{s}^m,\pmb{\lambda})\sqb{\bar{F}_{\mathrm{g}}(\norms{\bm{s}^m}_1)-\bar{F}^{\mathcal G}_{\mathrm{g}}(\norms{\bm{s}^m}_1)}}\IEEEnonumber\\
	&\overset{\text{(b)}}{\leq}&\sum_{m=1}^{M}\rho_m(\bm{s}^m,\pmb{\lambda})\abs{\bar{F}_{\mathrm{g}}(\norms{\bm{s}^m}_1)-\bar{F}^{\mathcal G}_{\mathrm{g}}(\norms{\bm{s}^m}_1)}\IEEEnonumber\\
	&\overset{\text{(c)}}{\leq}&\rb{4\sqrt{\frac{(2N+1)\ln(N+1)}{3N(N+1)}}+\sqrt{\frac{2\ln(2/\delta)}{N}}}\sum_{m=1}^{M}\rho_m(\bm{s}^m,\pmb{\lambda})\IEEEnonumber\\
	&\overset{\text{(d)}}{\leq}&\rb{4\sqrt{\frac{(2N+1)\ln(N+1)}{3N(N+1)}}+\sqrt{\frac{2\ln(2/\delta)}{N}}}\log_2(1+SP),
\end{IEEEeqnarray}
where (a) follows from \eqref{excpected_rate} and \eqref{empirical_excpected_rate}; (b) follows from from the triangle inequality and since the rate of each layer is non-negative; (c) follows from \cref{prop:CCDF_convergence}; and (d) holds since $S\geq \norms{\bm{s}}_1$.
Finally, based on inequalities \eqref{eq:basic_gen_ineq} and \eqref{eq:rate_conv}, we can upper bound the optimality gap as \eqref{eq:bound_opt_gap_1}.

\subsection{Proof of Proposition \ref{prop:CCDF_convergence}}\label{app:proof_ccdf_conv}
Let function $\ell:\mathbb R\times\mathbb C\mapsto \{0,1\}$ be defined as
\begin{IEEEeqnarray}{c}\label{eq:def_ell}
	\ell(s,g)\triangleq \mathbf{1}_{g\geq s}.
\end{IEEEeqnarray}
The true and empirical \ac{CCDF} can hence be expressed as
\begin{IEEEeqnarray}{c}
	\bar{F}_{\mathrm{g}}(s)=\mathbb E_{\mathrm g}\sqb{\ell(s,\mathrm{g})}
\end{IEEEeqnarray}
and
\begin{IEEEeqnarray}{c}
	\bar{F}^{\mathcal G}_{\mathrm{g}}(s)=\frac{1}{N}\sum_{i=1}^{N}\ell(s,g_i),
\end{IEEEeqnarray}
respectively, where $g_1,\ldots,g_N\in\mathcal G$ are $N$ fading realizations. In addition, let $\mathcal L(g_1,\ldots,g_N)\subset \{0,1\}^N$ be the set
\begin{IEEEeqnarray}{c}\label{eq:def_set_L}
	\mathcal{L}(g_1,\ldots,g_N)\triangleq\{\rb{\ell(s,g_1),\ldots,\ell(s,g_N)}:s\in\mathbb R\}.
\end{IEEEeqnarray}
Furthermore, denote by $\text{Rad}(\mathcal L(g_1,\ldots,g_N))$ the \emph{Rademacher complexity} of set $\mathcal L(g_1,\ldots,g_N)$, i.e.,
\begin{IEEEeqnarray}{c}
	\text{Rad}(\mathcal L(g_1,\ldots,g_N))\triangleq \frac{1}{N}\mathbb E_{\mathbf{b}}\sqb{\sup_{\bm{\ell}\in\mathcal L(g_1,\ldots,g_N)}\sum_{i=1}^{N}\mathrm{b}_i\ell_i},
\end{IEEEeqnarray}
where the elements of random vector $\mathbf{b}=(\mathrm{b}_1,\ldots,\mathrm{b}_N)\in\{\pm 1\}^N$ are \ac{iid} with $\Pr[b_i=1]=\Pr[b_i=-1]=1/2$. Since $|\ell(s,g)|\leq 1$ for all $g\in\mathbb R_+$ and $s\in\mathbb R$, by \cite[Thm 26.5]{shalev2014understanding} and \cite[Prop. 8]{weinberger2021generalization}, for random variables $\mathrm{g}_1,\ldots,\mathrm{g}_N$ that are \ac{iid} according to $p_{\mathrm{g}}(g)$, we have
\begin{IEEEeqnarray}{c}\label{eq:T26.5}
	\Pr\sqb{\bigcap_{s\in\mathbb R}\cb{\abs{\bar{F}_{\mathrm{g}}(s)-\bar{F}^{\mathcal G}_{\mathrm{g}}(s)}\leq 4\mathbb E\sqb{\text{Rad}(\mathcal L(\mathrm{g}_1,\ldots,\mathrm{g}_N))}+\sqrt{\frac{2\ln(2/\delta)}{N}}}}\geq 1-\delta.
\end{IEEEeqnarray}

Next, we bound the expected Rademacher complexity $\mathbb E\sqb{\text{Rad}(\mathcal L(\mathrm{g}_1,\ldots,\mathrm{g}_N))}$ in \eqref{eq:T26.5}. We assume, \ac{WLOG}, that the channel realizations $g_1,\ldots,g_N\in\mathcal G$ are ordered such that $g_i\geq g_j$ for all $j\in[i]$. Note that, if $\ell(s,g_j)=1$ for some $s\in\mathbb R$ then $\ell(s,g_i)=1$ for all $j\leq i\leq N$. Therefore, we have
\begin{IEEEeqnarray}{c}
	|\mathcal{L}(g_1,\ldots,g_N)|=N+1.
\end{IEEEeqnarray}
Denote by
\begin{IEEEeqnarray}{c}
	\bar{\bm{\ell}}\triangleq\frac{1}{N+1}\sum_{\bm{\ell}\in\mathcal{L}(g_1,\ldots,g_N)}\bm{\ell}=\frac{1}{N+1}(1,2,\ldots,N)
\end{IEEEeqnarray}
the average vector in $\mathcal{L}(g_1,\ldots,g_N)$. Note that
\begin{IEEEeqnarray}{c}
	\max_{\bm{\ell}\in\mathcal{L}(g_1,\ldots,g_N)}\norm{\bm{\ell}-\bar{\bm{\ell}}}_2=\norm{\bar{\bm{\ell}}}_2=\sqrt{\frac{N(2N+1)}{6(N+1)}}.
\end{IEEEeqnarray}
Hence, by \emph{Massart Lemma} \cite[Lemma 26.8]{shalev2014understanding}, we have 
\begin{IEEEeqnarray}{c}\label{app:eq_rademacher_ub}
	\text{Rad}(\mathcal L(g_1,\ldots,g_N))\leq \sqrt{\frac{N(2N+1)}{6(N+1)}}\cdot\frac{\sqrt{2\ln(N+1)}}{N}=\sqrt{\frac{(2N+1)\ln(N+1)}{3N(N+1)}}
\end{IEEEeqnarray}
for any channel realizations $g_1,\ldots,g_N\in\mathbb R_+$. This implies that the upper bound in \eqref{app:eq_rademacher_ub} bounds the expected Rademacher complexity $\mathbb E\sqb{\text{Rad}(\mathcal L(\mathrm{g}_1,\ldots,\mathrm{g}_N))}$ as well. By substituting \eqref{app:eq_rademacher_ub} in \eqref{eq:T26.5} we get \eqref{eq:prop_convergence}.

\subsection{Proof of Proposition \ref{prop:opt_r}}\label{app:proof_opt_r}
First, we prove that $f^{\mathcal G}_\beta(\bm{s},\pmb{\lambda},r)$ is concave with respect to $r\geq 0$. To this end, let $0\leq r_1\leq r_2$, $\delta\in[0,1]$, and assume that, \ac{WLOG}, dataset $\mathcal{G}$ is sorted such that $g_1\leq g_2\leq\cdots\leq g_N$. Therefore, there exist $0\leq N_1\leq N_2$ such that we can express $f^{\mathcal G}_\beta(\bm{s},\pmb{\lambda},r_1)$ and $f^{\mathcal G}_\beta(\bm{s},\pmb{\lambda},r_2)$ as
\begin{IEEEeqnarray}{c}
	f^{\mathcal G}_\beta(\bm{s},\pmb{\lambda},r_1)=r_1-\frac{1}{N\beta}\sum_{i=1}^{N_1}\rb{r_1-R(\bm{s},\pmb{\lambda},g_i)}
\end{IEEEeqnarray}
and
\begin{IEEEeqnarray}{c}
	f^{\mathcal G}_\beta(\bm{s},\pmb{\lambda},r_2)=r_2-\frac{1}{N\beta}\sum_{i=1}^{N_2}\rb{r_2-R(\bm{s},\pmb{\lambda},g_i)},
\end{IEEEeqnarray}
respectively. Furthermore, there exists $N_\delta$, where $N_1\leq N_\delta \leq N_2$, such that we have
\begin{IEEEeqnarray}{c}
	f^{\mathcal G}_\beta(\bm{s},\pmb{\lambda},\delta r_1+(1-\delta)r_2)=\delta r_1+(1-\delta)r_2-\frac{1}{N\beta}\sum_{i=1}^{N_\delta}\rb{\delta r_1+(1-\delta)r_2-R(\bm{s},\pmb{\lambda},g_i)}.
\end{IEEEeqnarray}
Note that we can lower bound $f^{\mathcal G}_\beta(\bm{s},\pmb{\lambda},\delta r_1+(1-\delta)r_2)$ as
\begin{IEEEeqnarray}{rCl}
	f^{\mathcal G}_\beta(\bm{s},\pmb{\lambda},\delta r_1+(1-\delta)r_2)&=& \delta\sqb{r_1-\frac{1}{N\beta}\sum_{i=1}^{N_\delta}\rb{ r_1-R(\bm{s},\pmb{\lambda},g_i)}}\IEEEnonumber\\
	&&+(1-\delta)\sqb{r_2-\frac{1}{N\beta}\sum_{i=1}^{N_\delta}\rb{ r_2-R(\bm{s},\pmb{\lambda},g_i)}}\IEEEnonumber\\
	&=& \delta\sqb{f^{\mathcal G}_\beta(\bm{s},\pmb{\lambda},r_1)-\frac{1}{N\beta}\sum_{i=N_1+1}^{N_\delta}\rb{ r_1-R(\bm{s},\pmb{\lambda},g_i)}}\IEEEnonumber\\
	&&+(1-\delta)\sqb{f^{\mathcal G}_\beta(\bm{s},\pmb{\lambda},r_2)+\frac{1}{N\beta}\sum_{i=N_\delta+1}^{N_2}\rb{ r_2-R(\bm{s},\pmb{\lambda},g_i)}}\IEEEnonumber\\
	&\geq& \delta f^{\mathcal G}_\beta(\bm{s},\pmb{\lambda},r_1)+(1-\delta)f^{\mathcal G}_\beta(\bm{s},\pmb{\lambda},r_2),
\end{IEEEeqnarray}
where the inequality holds since $r_1<R(\bm{s},\pmb{\lambda},g_i)$ for $i>N_1$, and $r_2\geq R(\bm{s},\pmb{\lambda},g_i)$ for $i\leq N_2$. Therefore, $f^{\mathcal G}_\beta(\bm{s},\pmb{\lambda},r)$ is concave with respect to $r\geq 0$. 

Similarly, for $r\geq 0$, let $N_r\geq 0$ such that
\begin{IEEEeqnarray}{c}
	f^{\mathcal G}_\beta(\bm{s},\pmb{\lambda},r)=r-\frac{1}{N\beta}\sum_{i=1}^{N_r}\rb{r-R(\bm{s},\pmb{\lambda},g_i)}.
\end{IEEEeqnarray}
The derivative of $f^{\mathcal G}_\beta(\bm{s},\pmb{\lambda},r)$ with respect to $r$ is hence given as
\begin{IEEEeqnarray}{c}\label{eq:d_dr_f}
	\frac{\partial}{\partial r}f^{\mathcal G}_\beta(\bm{s},\pmb{\lambda},r) = 1-\frac{N_r}{N\beta}.
\end{IEEEeqnarray}
We find $r^{\mathcal G}_\beta(\bm{s},\pmb{\lambda})$ by analyzing the case of $N\beta>1$ and the case of $N\beta\leq 1$ separately.

\subsubsection{$N\beta>1$}
If $N\beta$ is an integer, the derivative in \eqref{eq:d_dr_f} is zero iff $N_r=N\beta$. Therefore, the optimal $r$ is given as $r^{\mathcal G}_\beta(\bm{s},\pmb{\lambda})=R(\bm{s},\pmb{\lambda},g_{N\beta})$. If $N\beta$ is not an integer, the optimal $r$ is that which induces $N_r$ closest to $N\beta$, i.e., $r^{\mathcal G}_\beta(\bm{s},\pmb{\lambda})=R(\bm{s},\pmb{\lambda},g_{\nint{N\beta}})$.

\subsubsection{$N\beta\leq 1$}
In this case, the derivative in \eqref{eq:d_dr_f} cannot be zero and is positive iff $N_r=0$. Therefore, the optimal $r$ is the largest $r$ for which $f^{\mathcal G}_\beta(\bm{s},\pmb{\lambda},r)=r$, i.e., $r^{\mathcal G}_\beta(\bm{s},\pmb{\lambda})=R(\bm{s},\pmb{\lambda},g_{1})$. Note that, for $N\beta\leq 1$, we have $\nint{N\beta}=1$ since we defined $\nint{x}$ to be the closest \emph{positive} integer to scalar $x$.

\subsection{Proof of Proposition \ref{prop:f_convergence}}\label{app:proof_prop_f_conv}
As detailed in \cref{sec:CVaR}, the \ac{BS} optimizes the $\beta$-\ac{CVaR} rate $R_\beta(\bm{s},\pmb{\lambda})$ by optimizing the function $f_\beta(\bm{s},\pmb{\lambda},r)$ in \eqref{eq:def_f}. Therefore, the optimality gap $e_\beta^{\mathcal G}$ \eqref{eq:gen_error_beta} can be restated as
\begin{IEEEeqnarray}{c}\label{eq:opt_gap_restated}
	e_\beta^{\mathcal G}=f_{\beta}\big(\bm{s}_\beta^*,\pmb{\lambda}_\beta^*,r_\beta(\bm{s}_\beta^*,\pmb{\lambda}_\beta^*)\big)-f_{\beta}\big(\bm{s}_\beta^{\mathcal G},\pmb{\lambda}_\beta^{\mathcal G},r_\beta^{\mathcal G}(\bm{s}_\beta^{\mathcal G},\pmb{\lambda}_\beta^{\mathcal G})\big).
\end{IEEEeqnarray}
Next, we bound the optimality gap \eqref{eq:opt_gap_restated} as
\begin{IEEEeqnarray}{rCl}\label{eq:basic_gen_ineq_beta}
	e_\beta^{\mathcal G}&=&f_\beta\big(\bm{s}_\beta^*,\pmb{\lambda}_\beta^*,r_\beta(\bm{s}_\beta^*,\pmb{\lambda}_\beta^*)\big)-f^{\mathcal G}_{\beta}\big(\bm{s}_\beta^*,\pmb{\lambda}_\beta^*,r_\beta(\bm{s}_\beta^*,\pmb{\lambda}_\beta^*)\big)\IEEEnonumber\\
	&&+f^{\mathcal G}_{\beta}\big(\bm{s}_\beta^*,\pmb{\lambda}_\beta^*,r_\beta(\bm{s}_\beta^*,\pmb{\lambda}_\beta^*)\big)-f^{\mathcal G}_{\beta}\big(\bm{s}_\beta^{\mathcal G},\pmb{\lambda}_\beta^{\mathcal G},r_\beta^{\mathcal G}(\bm{s}_\beta^{\mathcal G},\pmb{\lambda}_\beta^{\mathcal G})\big)\IEEEnonumber\\
	&&+f^{\mathcal G}_{\beta}\big(\bm{s}_\beta^{\mathcal G},\pmb{\lambda}_\beta^{\mathcal G},r_\beta^{\mathcal G}(\bm{s}_\beta^{\mathcal G},\pmb{\lambda}_\beta^{\mathcal G})\big)-f_\beta\big(\bm{s}_\beta^{\mathcal G},\pmb{\lambda}_\beta^{\mathcal G},r_\beta^{\mathcal G}(\bm{s}_\beta^{\mathcal G},\pmb{\lambda}_\beta^{\mathcal G})\big)\IEEEnonumber\\
	&\leq&\rb{f_\beta\big(\bm{s}_\beta^*,\pmb{\lambda}_\beta^*,r_\beta(\bm{s}_\beta^*,\pmb{\lambda}_\beta^*)\big)-f^{\mathcal G}_{\beta}\big(\bm{s}_\beta^*,\pmb{\lambda}_\beta^*,r_\beta(\bm{s}_\beta^*,\pmb{\lambda}_\beta^*)\big)}\IEEEnonumber\\
	&&+\rb{f^{\mathcal G}_{\beta}\big(\bm{s}_\beta^{\mathcal G},\pmb{\lambda}_\beta^{\mathcal G},r_\beta^{\mathcal G}(\bm{s}_\beta^{\mathcal G},\pmb{\lambda}_\beta^{\mathcal G})\big)-f_\beta\big(\bm{s}_\beta^{\mathcal G},\pmb{\lambda}_\beta^{\mathcal G},r_\beta^{\mathcal G}(\bm{s}_\beta^{\mathcal G},\pmb{\lambda}_\beta^{\mathcal G})\big)},
\end{IEEEeqnarray}
where the inequality holds since $(\bm{s}_\beta^{\mathcal G},\pmb{\lambda}_\beta^{\mathcal G},r_\beta^{\mathcal G}(\bm{s}_\beta^{\mathcal G},\pmb{\lambda}_\beta^{\mathcal G}))$ maximize the empirical approximation $f^{\mathcal G}_{\beta}(\bm{s},\pmb{\lambda},r)$ \eqref{eq:ampir_f}. In addition, note that, for rate allocation vectors with bounded norms $\normm{\bm{s}}_1\leq S$, we have $r_\beta(\bm{s},\pmb{\lambda}),r_\beta^{\mathcal{G}}(\bm{s},\pmb{\lambda})\leq \log_2(1+SP)\triangleq \tilde{R}$. 

Similar to the proof of \cref{prop:expected_rate_convergence}, to further bound the optimality gap, we bound, uniformly, the difference $|f_\beta(\bm{s},\pmb{\lambda},r)-f^{\mathcal G}_\beta(\bm{s},\pmb{\lambda},r)|$ for all $\pmb{\lambda}\in\Delta_c^M$, $0\leq r\leq\tilde{R}$ and $\bm{s}\in\mathbb R_+^M$ with $\norms{\bm{s}}_1\leq S$.
Note that the difference $|f_\beta(\bm{s},\pmb{\lambda},r)-f^{\mathcal G}_\beta(\bm{s},\pmb{\lambda},r)|$ can be expressed as
\begin{IEEEeqnarray}{c}
	\abs{f_\beta(\bm{s},\pmb{\lambda},r)-f^{\mathcal G}_\beta(\bm{s},\pmb{\lambda},r)}=\beta^{-1}\abs{\frac{1}{N}\sum_{i=1}^{N}\delta^+(\bm{s},\pmb{\lambda},g_i,r)-\mathbb E_{\mathrm g}\sqb{\delta^+(\bm{s},\pmb{\lambda},\mathrm{g},r)}},
\end{IEEEeqnarray}
where we have defined function $\delta^+(\bm{s},\pmb{\lambda},g,r)\triangleq\rb{r-R(\bm{s},\pmb{\lambda},g)}^{+}$.
Let $\mathcal{D}^+(g_1,\ldots,g_N)\subset\mathbb R_+^N$ be the set
\begin{IEEEeqnarray}{ll}
	\mathcal{D}^+(g_1,\ldots,g_N)\triangleq &\big\{\rb{\delta^+(\bm{s},\pmb{\lambda},g_1,r),\ldots,\delta^+(\bm{s},\pmb{\lambda},g_N,r)}:\IEEEnonumber\\
	&\quad\bm{s}\in\mathbb R_+^M,~\norms{\bm{s}}_1\leq S,~\pmb{\lambda}\in\Delta_c^M,~0\leq r\leq \tilde{R}\big\}.
\end{IEEEeqnarray}
Since $|\delta^+(\bm{s},\pmb{\lambda},g,r)|\leq\tilde{R}$ for all $\bm{s}\in\mathbb R_+^M$, $\pmb{\lambda}\in\Delta_c^M$, $g\in\mathbb R_+$, and $0\leq r\leq \tilde{R}$, by \cite[Thm 26.5]{shalev2014understanding} and \cite[Prop. 8]{weinberger2021generalization}, for random variables $\mathrm{g}_1,\ldots,\mathrm{g}_N$ that are \ac{iid} according to $p_{\mathrm{g}}(g)$, we have, with probability at least $1-\delta$,
\begin{IEEEeqnarray}{c}\label{app:eq_bound_conv_f}
	\abs{f_\beta(\bm{s},\pmb{\lambda},r)-f^{\mathcal G}_\beta(\bm{s},\pmb{\lambda},r)}\leq \beta^{-1}\rb{ 4\mathbb E\sqb{\text{Rad}(\mathcal{D}^+(\mathrm{g}_1,\ldots,\mathrm{g}_N))}+\tilde{R}\sqrt{\frac{2\ln(2/\delta)}{N}}}.
\end{IEEEeqnarray}

Next, we bound the Rademacher complexity $\text{Rad}(\mathcal{D}^+(\mathrm{g}_1,\ldots,\mathrm{g}_N))$ in \eqref{app:eq_bound_conv_f}. Let $\mathcal{D}(g_1,\ldots,g_N)$ be the set
\begin{IEEEeqnarray}{ll}
	\mathcal{D}(g_1,\ldots,g_N)\triangleq &\big\{\rb{\delta(\bm{s},\pmb{\lambda},g_1,r),\ldots,\delta(\bm{s},\pmb{\lambda},g_N,r)}:\IEEEnonumber\\
	&\quad\bm{s}\in\mathbb R_+^M,~\norms{\bm{s}}_1\leq S,~\pmb{\lambda}\in\Delta_c^M,~0\leq r\leq \tilde{R}\big\},
\end{IEEEeqnarray}
where we have defined function 
\begin{IEEEeqnarray}{c}\label{eq:def_delta}
	\delta(\bm{s},\pmb{\lambda},g,r)\triangleq r-R(\bm{s},\pmb{\lambda},g).
\end{IEEEeqnarray}
Note that the set $\mathcal{D}^+(g_1,\ldots,g_N)$ can be expressed as
\begin{IEEEeqnarray}{ll}
	\mathcal{D}^+(g_1,\ldots,g_N)=&\big\{\rb{q(\delta(\bm{s},\pmb{\lambda},g_1,r)),\ldots,q(\delta(\bm{s},\pmb{\lambda},g_N,r))}:\IEEEnonumber\\
	&\quad\bm{s}\in\mathbb R_+^M,~\norms{\bm{s}}_1\leq S,~\pmb{\lambda}\in\Delta_c^M,~0\leq r\leq \tilde{R}\big\},
\end{IEEEeqnarray}
where function $q:\mathbb R\mapsto\mathbb R_+$ is defined as $q(x)\triangleq (x)^+$. Since $q(\cdot)$ is a 1-Lipschitz function, i.e., for all $x,y\in\mathbb R$, we have $|q(x)-q(y)|\leq|x-y|$, then, by the \emph{Contraction Lemma} \cite[Lemma 26.9]{shalev2014understanding}, we have
\begin{IEEEeqnarray}{c}
	\text{Rad}(\mathcal{D}^+(g_1,\ldots,g_N))\leq \text{Rad}(\mathcal{D}(g_1,\ldots,g_N)).
\end{IEEEeqnarray}
In addition, let $\mathcal{R}(g_1,\ldots,g_N)\subset\mathbb R_+^N$ be the set
\begin{IEEEeqnarray}{ll}
	\mathcal{R}(g_1,\ldots,g_N)\triangleq &\big\{\rb{R(\bm{s},\pmb{\lambda},g_1),\ldots,R(\bm{s},\pmb{\lambda},g_N)}:
	\bm{s}\in\mathbb R_+^M,~\norms{\bm{s}}_1\leq S,~\pmb{\lambda}\in\Delta_c^M\big\}.
\end{IEEEeqnarray}
It follows from the definition of $\delta(\bm{s},\pmb{\lambda},g,r)$ in \eqref{eq:def_delta} and from \cite[Lemma 26.6]{shalev2014understanding} that
\begin{IEEEeqnarray}{c}
	\text{Rad}(\mathcal{D}(g_1,\ldots,g_N))\leq\text{Rad}(\mathcal{R}(g_1,\ldots,g_N)).
\end{IEEEeqnarray}
Note that the achievable rate $R(\bm{s},\pmb{\lambda},g)$ can be expressed as
\begin{IEEEeqnarray}{c}
	R(\bm{s},\pmb{\lambda},g)=\log_2(1+SP)\sum_{m=1}^{M}\alpha_m(\bm{s}^m,\pmb{\lambda})\ell(\norms{\bm{s}^m}_1,g),
\end{IEEEeqnarray}
where we have defined functions
\begin{IEEEeqnarray}{c}
	\alpha_m(\bm{s}^m,\pmb{\lambda})\triangleq \frac{\rho_m(\bm{s}^m,\pmb{\lambda})}{\log_2(1+SP)},\quad m\in[M],
\end{IEEEeqnarray}
and function $\ell(\cdot,\cdot)$ is defined in \eqref{eq:def_ell}. Now, let  $\tilde{\mathcal{R}}(g_1,\ldots,g_N)\subset\mathbb R_+^N$ be the set
\begin{IEEEeqnarray}{ll}
	\tilde{\mathcal{R}}(g_1,\ldots,g_N)\triangleq &\Bigg\{\sum_{m=1}^{M}\alpha_m(\bm{s}^m,\pmb{\lambda})\rb{\ell(\norms{\bm{s}^m}_1,g_1),\ldots,\ell(\norms{\bm{s}^m}_1,g_N)}:\IEEEnonumber\\
	&\quad\bm{s}\in\mathbb R_+^M,~\norms{\bm{s}}_1\leq S,~\pmb{\lambda}\in\Delta_c^M\Bigg\}.
\end{IEEEeqnarray}
It follows from \cite[Lemma 26.6]{shalev2014understanding} that
\begin{IEEEeqnarray}{c}
	\text{Rad}(\mathcal{R}(g_1,\ldots,g_N))\leq \log_2(1+SP)\text{Rad}(\tilde{\mathcal{R}}(g_1,\ldots,g_N)).
\end{IEEEeqnarray}
Furthermore, since $\alpha_m(\bm{s}^m,\pmb{\lambda})\geq0$ and $\sum_{m=1}^{M}\alpha_m(\bm{s}^m,\pmb{\lambda})\leq 1$, then $\text{Rad}(\tilde{\mathcal{R}}(g_1,\ldots,g_N))\subseteq \text{conv}(\mathcal{L}(g_1,\ldots,g_N))$, where set $\mathcal{L}(g_1,\ldots,g_N)$ is defined in \eqref{eq:def_set_L}. Hence, we have
\begin{IEEEeqnarray}{rCl}
	\text{Rad}(\tilde{\mathcal{R}}(g_1,\ldots,g_N))&\leq& \text{Rad}(\text{conv}(\mathcal{L}(g_1,\ldots,g_N)))\IEEEnonumber\\
	&\overset{\text{(a)}}{=}&\text{Rad}(\mathcal{L}(g_1,\ldots,g_N))\IEEEnonumber\\
	&\overset{\text{(b)}}{\leq}&\sqrt{\frac{(2N+1)\ln(N+1)}{3N(N+1)}},
\end{IEEEeqnarray}
where $(a)$ follows from \cite[Lemma 26.7]{shalev2014understanding} and (b) follows from \eqref{app:eq_rademacher_ub}. Overall, we have
\begin{IEEEeqnarray}{c}\label{eq:app_rad_D_bound}
	\text{Rad}(\mathcal{D}^+(\mathrm{g}_1,\ldots,\mathrm{g}_N))\leq \log_2(1+SP)\sqrt{\frac{(2N+1)\ln(N+1)}{3N(N+1)}}.
\end{IEEEeqnarray}
Substituting \eqref{eq:app_rad_D_bound} in \eqref{app:eq_bound_conv_f} yields 
\begin{IEEEeqnarray}{rCl}\label{eq:f_conv}
	\abs{f_\beta(\bm{s},\pmb{\lambda},r)-f^{\mathcal G}_\beta(\bm{s},\pmb{\lambda},r)}&\leq& \beta^{-1}\Bigg(4\sqrt{\frac{(2N+1)\ln(N+1)}{3N(N+1)}}+\sqrt{\frac{2\ln(2/\delta)}{N}}\Bigg)\log_2(1+SP).\IEEEeqnarraynumspace
\end{IEEEeqnarray}
Finally, based on inequalities \eqref{eq:basic_gen_ineq_beta} and \eqref{eq:f_conv}, we can upper bound the optimality gap as \eqref{eq:bound_opt_gap_beta}.

\bibliographystyle{IEEEtran}
\bibliography{IEEEabrv,myBib}

\end{document}